\documentclass[cite]{cimento}

\usepackage{amsmath}
\usepackage{amsfonts}
\usepackage{bbm}
\usepackage{bm}
\usepackage{graphicx}
\usepackage{verbatim}
\usepackage{color}
\usepackage{amssymb}
\usepackage{latexsym}
\usepackage{wasysym}

\usepackage{epstopdf}

\def\bd{\boldsymbol}

\title{Edge physics in two-dimensional topological insulators}
\author{G.~Dolcetto\from{ins:x}\from{ins:z}\thanks{e-mail: giacomo.dolcetto@uni.lu}\ETC,
M.~Sassetti\from{ins:x}\from{ins:y}
\atque
T.~L.~Schmidt\from{ins:z}}
\instlist{
\inst{ins:x}  SPIN-CNR, Via Dodecaneso 33, 16146, Genova, Italy.
\inst{ins:y}  Dipartimento di Fisica, Universit\`{a} di Genova, Via Dodecaneso 33, 16146, Genova, Italy.
\inst{ins:z}  Physics and Materials Science Research Unit, University of Luxembourg, L-1511 Luxembourg}
\PACSes{
\PACSit{73.23.-b}{Electronic transport in mesoscopic systems}
\PACSit{73.21.-b}{Electron states and collective excitations in multilayers, quantum wells, mesoscopic, and nanoscale systems}
\PACSit{72.10.-d}{Theory of electronic transport; scattering mechanisms}
\PACSit{73.43.-f}{Quantum Hall effects}
}

\begin{document}

\maketitle

\begin{abstract}
Topology in condensed matter physics manifests itself in the emergence of edge or surface states protected by underlying symmetries.
We review two-dimensional topological insulators whose one-dimensional edge states are characterized by spin-momentum locking and protected by time-reversal symmetry.
We focus in particular on their transport properties in the presence of electron interactions, which can allow the onset of different backscattering mechanisms, thus leading to deviations from the quantized conductance observed in the ballistic regime.
The combined presence of helicity and electron interactions creates a new paradigm of the one-dimensional world called helical Luttinger liquid, whose theoretical properties and experimental observations are reviewed.
\end{abstract}

\tableofcontents

\section{Introduction: The era of topological condensed matter physics}
\label{intro}
Solid state, particle, and mathematical physics represent three fundamental branches of modern physics.
In the last decades, merging ideas coming from these branches has led to a new paradigm, the era of topological condensed matter physics.
The notion of band structure, typical of solid state physics, together with topological concepts, connected to the field of mathematical physics, allows to provide a topological classification of materials.
The result is the emergence of topologically non-trivial electronic systems described by Dirac-like equations, which are naturally found in particle and nuclear physics~\cite{shen2013topological}.
Even though it is known that a material can be classified, according to its band structure, as an insulator or a metal, with large or low resistance respectively, in 2007 a material which does not fit in this classification was discovered: the first topological insulator~\cite{konig2007quantum, hasan2010topological}.

To be more precise, the advent of topological condensed matter physics can be dated back to 1980, when von Klitzing \textit{et al.} discovered experimentally~\cite{von1980new} that when a two-dimensional (2d) electron gas at the interface of a semiconductor heterojunction is subjected to a strong magnetic field, the longitudinal conductance vanishes whereas the Hall conductance is quantized to the values $n e^2/h$, where $n$ is an integer number.
This integer quantum Hall (QH) effect was explained~\cite{laughlin1981quantized} within a single-particle picture in terms of Landau levels, and it was realized that these systems host metallic edge states coexisting with insulating bulk states.
Semiclassically, this effect can be understood by imagining the electrons in the bulk of the 2d plane to be localised due to their cyclotron motion, while the electrons near the edges perform skipping orbits, thus creating chiral one-dimensional (1d) conducting channels~\cite{Halperin1982quantized}.
Surprisingly, the measured quantised conductance is almost completely insensitive to the presence of impurities and disorder. The explanation relies on the topological nature of the QH states, which form a quantum phase which cannot be described according to the usual Ginzburg-Landau theory.
This classification, which is based on local order parameters, fails to characterise QH states. Instead, topological quantum numbers must be introduced which reflect the global properties of the system.
The most striking manifestation of non-trivial topology is the emergence of edge states: in QH systems with Hall conductance $n e^2/h$, $n$ channels are located at the sample boundaries.
The insensitivity of the QH conductance to fabrication dependent details is a consequence of the topological properties of the band structure. Since the latter are not affected by weak local perturbations such as disorder and impurities, the metallic edge states are robust.
The QH effect at filling factor $n$ is classified as $\mathbb{Z}_n$ topological order, i.e., the phase is characterized by an integer number $n$, which determines the quantised conductance $\sigma_{H}=n e^2/h$ and the presence of $n$ edge channels.

Soon after the discovery of the QH effect, physicists started to wonder if time-reversal symmetry (TRS) breaking, which in QH systems is brought about by the magnetic field, is necessary to generate a topologically non-trivial state. Related to this question, Haldane proposed a lattice system of spinless electrons in a periodic magnetic flux to realise the QH effect~\cite{Haldane1988model}. Although the total magnetic flux through the system is zero, electrons form a conducting edge channel. As no net magnetic field is present, the quantised Hall conductance in this case must originate from the band structure of electrons rather than from the discrete Landau levels in a magnetic field. Several years later, a breakthrough was made by Kane and Mele~\cite{kane2005quantum, kane2005z} who generalised the Haldane model to a graphene sheet with spin-orbit coupling (SOC), providing the theoretical basis for 2d topological insulators (TIs). SOC respects TRS and it acts as two effective and opposite magnetic fields for spin-up and spin-down electrons. As in Haldane's proposal, metallic edge states appear in the bulk energy gap. However, due to TRS, they now come in Kramers pairs, with electrons with opposite spin propagating in opposite directions.
Crucially, due to Kramers' theorem, these edge states are protected from backscattering (BS) as long as TRS is preserved, thus giving rise to a symmetry-protected topological phase.
These edge stares are called \emph{helical}, because the electron spin is locked to the propagation direction in such a way that they are eigenstates of the helicity operators, i.e., the projection of the spin onto the momentum.
The edge states give rise to the quantum spin Hall (QSH) effect: a longitudinal charge current produces a transverse spin bias, resulting in a quantised spin Hall conductance $\sigma_{sH}=e/2\pi$.

Although it was soon realised that because of the tiny bulk gap, the QSH effect cannot be experimentally observed in graphene, the ideas emerging from these studies pushed theorists to search for the QSH effect in materials with strong intrinsic SOC.
A breakthrough was made in 2006 by Bernevig, Hughes, and Zhang~\cite{bernevig2006quantum} with the prediction that a CdTe/HgTe/CdTe quantum well (QW) behaves, under certain conditions, as a 2d TI.
The different band structures of HgTe and CdTe are responsible for a topological phase transition when the thickness of the HgTe layer of the QW is increased. CdTe has a normal semiconductor band structure, while HgTe has an inverted semimetallic one due to intrinsic SOC, with the $\Gamma_8$ band, usually originating from the valence band, above the $\Gamma_6$ band. If the HgTe layer is thicker than a critical thickness $d_c\approx 6.3$ nm, the QW displays an inverted band structure. Due to the closing and reopening of the gap between the $\Gamma_6$-like and $\Gamma_8$-like bands in passing from the normal to the inverted regime, a topological phase transition occurs, and helical edge states appear in the inverted regime.
This theoretical prediction was confirmed in 2007~\cite{konig2007quantum}, when non-local transport properties which could be ascribed to the helical edge states were observed.

Kane and Mele had already showed~\cite{kane2005z} that, contrary to the QH effect, the QSH effect is classified by a $\mathbb{Z}_2$ topological order.
If an even number of Kramers doublets appear on the edge, the edge state spectrum can be gapped out by various perturbations without breaking TRS, making such systems topologically equivalent to ordinary insulators. On the other hand, in the presence of an odd number of Kramers doublets a single metallic edge states is always present (as long as TRS is preserved), making such systems topologically nontrivial.
Therefore, the QSH conductance of a topological (trivial) insulator is $\sigma_{sH}=e/2\pi$ ($\sigma_{sH} = 0$), and its associated topological index is $n_s=1$ ($n_s = 0$).
Physically, the $\mathbb{Z}_2$ invariant $n_s$ simply counts the number of gapless and protected helical edge states modulo $2$, in analogy with the integer $n$ counting the number of edge channels in the QH.

In the meanwhile, the QSH phase has been theoretically predicted and experimentally observed in other materials as well.
A prominent example is an InAs/GaSb heterostructure.
An interface of these materials forms a type-II (staggered) junction, and the valence band of GaSb can hybridize with the conduction band of InAs.
In this case gate voltages are able to tune the system between the normal and inverted regime, allowing to investigate the topological phase transition with a great flexibility.
More closely related to the original graphene-like models proposed by Kane and Mele, it was found that a larger bulk gap could be achieved in silicene~\cite{liu2011quantum}, a lattice of silicon atoms~\cite{ezawa2012topological}.
Silicene, as well as other lattice models like germanene, stanene~\cite{ezawa2015monolayer} or transition-metal dichalcogenides~\cite{qian2014quantum, Nie2015quantum}, are promising for generating the QSH effect in lattice models.
We briefly mention the recent proposal of realizing topological insulators in arrays of tunnel coupled quantum wires, an interesting architecture which could give rise to several exotic states of matter~\cite{oreg2014fractional, sagi2014non, sagi2015fractional}.
The search for new 2d TIs, as well as a deeper understanding of the behaviour of the discovered ones, is currently one of the major issues in condensed matter physics.

Three dimensional (3d) TIs, characterised by insulating bulk states and metallic surface states with helical spin textures, have been discovered as well. Relevant examples are represented by Bi$_2$Se$_3$, Bi$_2$Te$_3$ and Sb$_2$Te$_3$~\cite{zhang2009topological,liu2010model}. As in the case of 2d TIs, the surface states are characterised by spin-momentum locking, with spin and momentum orthogonal one to each other. Angle-resolved photo-emission spectroscopy (ARPES) and spin-resolved ARPES detected the presence of surface states characterised by a Dirac cones with the predicted helical spin texture~\cite{xia2009observation,chen2009experimental,hsieh2009observation}.

By combining TIs with superconducting materials, a new building block for futuristic applications can be engeneered: the topological superconductor~\cite{qi2011topological}.
Here peculiar bound states appear, whose properties can be described in terms of Majorana fermions~\cite{kitaev2003fault}. These states have exotic properties, the most stunning being non-Abelian statistics, that in principle would allow to implement revolutionary protocols for fault-tolerant quantum computation~\cite{beenakker2011search, alicea2012new}.
Although the first theoretical models~\cite{oreg2010helical, lutchyn2010majorana} and experimental realisations~\cite{mourik2012signatures, das2012zero} do not exploit TIs, topological superconductors were theoretically proposed~\cite{Fu2008superconducting, Fu2009Josephson} and experimentally realised~\cite{hart2014induced, pribiag2014edge} in TI based platforms as well.
Although the final goal of these efforts is to implement topological superconductors, they also represent a valuable playground to shed light on the helical edge states and their physics.

The search for new topological materials does not end here.
For example, by driving ordinary insulators out of equilibrium by means of time-dependent perturbations a topological state can be achieved, called Floquet topological insulator~\cite{cayssol2013floquet}.
It was predicted that by irradiating an otherwise topologically-trivial HgTe/CdTe heterostructure with electromagnetic radiation, a topological phase transition can be achieved~\cite{lindner2011floquet}, with the subsequent emergence of protected edge states~\cite{torres2014multiterminal}.
Therefore, Floquet theory allows to extend the concept of a TI to out-of-equilibrium regimes~\cite{katan2013modulated}.

Other topologically nontrivial materials, which currently receive a lot of attention, include Weyl semimetals~\cite{Wang2012absence, balents2011viewpoint, burkov2011weyl, turner2013beyond}, topological crystalline insulators~\cite{fu2011topological, hsieh2012topological}, flat band models~\cite{sun2011nearly, sheng2011fractional} and fractional QH~\cite{Tsui1982two, laughlin1983anomalous, Jain1989composite, wen1991non} and QSH states~\cite{levin2009fractional, levin2012classification}.

In this review we will focus on 2d TIs, with special emphasis on the properties of their helical edge states.

In Sec.~\ref{sec:2dTI} we first provide a qualitative description of the edge state mechanism in systems described by Dirac-like Hamiltonians. We discuss the constraints necessary for the bulk insulator to acquire a topological phase with edge states.
Then we describe the emergence of helical edge states in more realistic heterostructures, and show that the topology of the underlying 2d system determines the topological protection of the edge states.

In Sec.~\ref{sec:exp} we review several important experiments involving 2d TIs, measuring non-local transport, imaging the helical edge states, and confirming their spin polarization. We briefly comment on new experiments which combine 2d TIs with superconductors. Although these experiments aim mainly at realizing topological superconductors, they provide valuable information about the physical properties of  the helical edge states.

In Sec.~\ref{sec:scattering} we discuss transport in edge states in more detail by considering the backscattering mechanisms occurring at the edge beyond the ballistic regime.
A crucial role is played by electron-electron interactions, which have a particularly strong impact on the physics of 1d systems.
Here the concept of single fermionic quasiparticles fails, and the relevant excitations have collective bosonic character.
The interacting helical edge states, described by the helical Luttinger liquid model, are expected to show a variety of unexpected physical properties.
The transport properties of the interacting helical edge states can be affected by peculiar backscattering mechanisms, which will be reviewed in this section.
Finally we discuss the recent experiment by Li \textit{et al.}~\cite{li2015observation} where evidence of helical Luttinger liquid behaviour was found.

In Sec.~\ref{sec:tunneling} we discuss the tunneling properties in 2d TIs. We introduce the quantum point contact geometry, showing that it can be useful both to investigate the fundamental physical properties of the helical edge states and to develop interesting devices in the context of electron quantum optics, such as electron interferometers.

Finally, Sec.~\ref{sec:outlook} briefly summarizes the main topics covered in this review.

\section{Two-dimensional topological insulators and their edge states}\label{sec:2dTI}
In order to provide a generic theoretical description of 2d TIs without limiting the discussion to a specific solid-state realisation, we start with the so-called modified Dirac Hamiltonian~\cite{shen2013topological}, which allows one to understand the main physical properties of 2d TIs and the emergence of helical edge states realising the QSH effect.

\subsection{Dirac equation in condensed matter systems and the emergence of bound states}\label{subsec:dirac}
The Dirac equation~\cite{dirac1928quantum} is known to describe an elementary relativistic spin-$1/2$ particle
\begin{equation}\label{eq1}
\bd{\mathcal{H}}=c p_i\bd{\alpha}_i+mc^2\bd{\beta}.
\end{equation}
Here, $c$ is the speed of light, $m$ is the rest mass of the particle, and $\bd{\alpha}_i$ and $\bd{\beta}$ are related to Dirac matrices satisfying the Clifford algebra.
The structure of the latter depends on the dimensionality $d$ of the system.
Before studying the case $d=2+1$, which is necessary to make connections with the physics of 2d TIs, it turns out to be useful to revisit the $d=1+1$ case. Indeed, many key concepts already emerge in this simpler case, and can then be easily generalised to higher dimensions. Furthermore, we replace the speed of light $c$ with a general Fermi velocity $v$.

We consider the 1d Dirac Hamiltonian density\footnote{The constant $\hbar$ is set to $1$ throughout this review, and explicitly restored for clarity when needed.}
\begin{equation}\label{eq2}
\bd{\mathcal{H}}=-i  v\partial_x\bd{\sigma}_x+M\bd{\sigma}_z
,\end{equation}
where we used $p_x=-i \partial_x$ and $M\equiv mv^2$.
Eigenstates with positive and negative energies are separated by the gap $\Delta=|2M|$.
In this case, the \emph{sign} of the energy gap is unimportant: the Dirac Hamiltonian is invariant under the transformation $M\to- M$, $\sigma_{y,z} \to -\sigma_{y,z}$ and $\sigma_x \to \sigma_x$, and so is the gap $\Delta$.
However, this aspect can be further investigated by looking at a system where the sign of the mass term changes.
Suppose that
\begin{equation}\label{m(x)}
M(x)=\left \{\begin{matrix} M_1<0 & \text{if }  x \ll 0 \\
M_2>0 & \text{if } x \gg 0 \end{matrix}\right..
\end{equation}
Finite energy gaps exist for $|x| \gg 0$. However, if we assume the mass $M(x)$ to interpolate continuously, the gap should vanish close to the domain wall at $x=0$, and close and reopen around it.
In a sense, the system is gapped almost everywhere, except near the domain wall, where a peculiar state with energy inside the bulk energy gap could exist.
Indeed, one finds that intra-gap bound states cannot exist if $\mathrm{sgn}(M_1M_2)=+1$. On the other hand, if $\mathrm{sgn}(M_1M_2)=-1$ a zero-energy bound state solution arises
\begin{equation}\label{eq:boundst}
\bd{\psi}(x)=\left (\begin{matrix}1 \\ i \end{matrix}\right )\sqrt{\frac{1}{v}\left |\frac{M_1M_2}{M_1-M_2}\right |}e^{-\frac{\left |M(x)x\right |}{v} },
\end{equation}
thus lying inside the bulk energy gap. Its probability distribution $|\bd{\psi}(x)|^2$ is localised around the domain wall and spreads over distances $\xi_{1(2)}= v/|M_{1(2)}|$.
Therefore, we conclude that the Dirac Hamiltonian admits intra-gap bound states localised near domain walls. In this sense, if we regard the vacuum as a system with an infinitely large and positive energy gap, a system with a negative gap would have intra-gap bound states at its boundaries.

Hence, opposite mass terms in Dirac Hamiltonians may explain the existence of edge states. However, due to the $M\to -M$ symmetry, one cannot distinguish bare Dirac Hamiltonians with opposite $M$ by a topological quantum number. This ambiguity is not present in the so-called modified Dirac Hamiltonian~\cite{shen2013topological}
\begin{equation}\label{Dirac_mod_px}
\bd{h}=v p_x\bd{\sigma}_x+\left (M+Bp_x^2\right )\bd{\sigma}_z,
\end{equation}
where the quadratic correction $Bp_x^2$ to the bare mass term has been added to break the symmetry $M\to -M$.
Translating to momentum space, one obtains
\begin{equation}\label{Dirac_mod}
\bd{h}(k_x)=v k_x\bd{\sigma}_x+M(k_x)\bd{\sigma}_z,
\end{equation}
with the $k_x$-dependent mass term
\begin{equation}\label{M(k)}
M(k_x)=M-B k_x^2
.\end{equation}
We are interested in studying whether Eq. \eqref{Dirac_mod_px} admits zero-energy bound states at its boundaries. Thus, we consider the Hamiltonian on a semi-infinite 1d system bounded at $x=0$, and require the wave function $\bd{\psi}(x)$ to vanish at $x=0$.
After multiplying by $\sigma_x$ from the left one can rewrite
\begin{equation}\label{Sch_mod}
\partial_x\bd{\psi}(x)=-\frac{1}{v }\left (M+B ^2\partial_x^2\right )\bd{\sigma}_y\bd{\psi}(x).
\end{equation}
Therefore, we can write $\bd{\psi}(x)=\bd{\chi}_{\eta}\phi(x)$, with $\phi(0)=0$ and $\bd{\chi}_{\eta}$ being an eigenstate of $\bd{\sigma}_y$ with eigenvalue $\eta = \pm 1$.
Equation \eqref{Sch_mod} now reduces to an equation for $\phi$
\begin{equation}
\partial_x\phi(x)=-\frac{\eta}{v }\left (M+B ^2\partial_x^2\right )\phi(x).
\end{equation}
Using the ansatz $\phi(x)=Ce^{-\lambda x}$, one obtains two solutions
\begin{equation}
\lambda_{\pm}=\frac{v}{2B}\left (\mathrm{sgn}(B)\pm\sqrt{1-4MB/v^2}\right ),
\end{equation}
that must satisfy $\lambda_{\pm}>0$ in order for $\bd{\psi}(x)$ to be normalisable.
This requirement leads to the criterion
\begin{equation}\label{eq:constrain}
\mathrm{sgn}(MB)=+1
,\end{equation}
which is the necessary condition for the existence of the bound state
\begin{equation}\label{eq:boundstate}
\bd{\psi}(x)\propto\frac{1}{\sqrt{2}}\left (\begin{matrix}\mathrm{sgn}(B) \\ i\end{matrix}\right )\left (e^{-x/\xi_+}-e^{-x/\xi_-}\right ).
\end{equation}
Here, $\xi_\pm^{-1}=v(1\pm\sqrt{1-4MB/v^2})/2|B|$ is related to the penetration depth of the edge state into the bulk.
\begin{figure}[!tt]
\centering
\includegraphics[width=7cm,keepaspectratio]{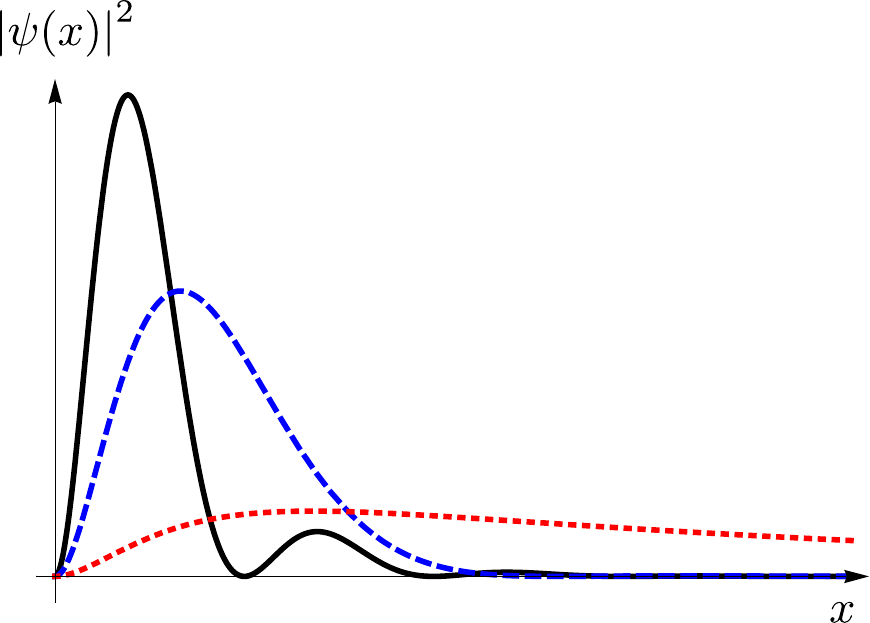}
\caption{Probability distribution of the bound state Eq. (\ref{eq:boundstate}): for large $MB>0$ the edge state is localised (solid black); by lowering $MB$ it spreads into the bulk (dashed blue), eventually disappearing for $MB\to 0$ (dotted red).}\label{OBC}
\end{figure}
The bound state is very sharply localised near the boundary for large $MB$, penetrates into the bulk by lowering $MB$, and eventually disappears when $MB\to 0$, as shown in Fig.~\ref{OBC}.
It is worth to note that the existence of the edge state only depends on the global quantity $\mathrm{sgn}(MB)$: if $\mathrm{sgn}(MB)=+1$ the edge state exists otherwise it does not, regardless of the specific values of the parameters $M$ and $B$.

\subsection{Generalisation to two dimensions}\label{subsec:generalisation}
In two spatial dimensions the modified Dirac Hamiltonian in momentum space reads ($k\equiv |\bd{k}|=k_x^2+k_y^2$)
\begin{eqnarray}\label{eq:Hchern}
\bd{h}\left (\bd{k}\right )=\bd{d}\left (\bd{k}\right )\cdot\bd{\sigma}
,\end{eqnarray}
with $\bd{\sigma}=(\bd{\sigma}_x,\bd{\sigma}_y,\bd{\sigma}_z)$,
\begin{eqnarray}\label{eq:vecd}
\bd{d}\left (\bd{k}\right )=\left (vk_x,-vk_y,M\left (k\right )\right )
,\end{eqnarray}
and the $k$-dependent mass term
\begin{equation}\label{eq:diracM(k)}
M\left(k\right)=M-B\left(k_x^2+k_y^2\right)
.\end{equation}
In order to study the emergence of edge states, we proceed similarly to the 1d case. We consider a semi-infinite plane ($x>0$), and require the solution to vanish at the boundary $x=0$. The system is translational invariant in the $\hat{y}$ direction so that $k_y$ is a good quantum number, but $k_x$ must be replaced by $-i\partial_x$.
Using the 1d solution, it is clear that a zero-energy bound state as in Eq.~\eqref{eq:boundstate} exists for $k_y=0$ if the constraint $\mathrm{sgn}(MB)=+1$ is satisfied. Then, the solution for non-zero $k_y$ corresponds to a bound state along $x$ and a plane wave along $y$
\begin{equation}\label{eq:boundstate2}
\bd{\psi}_{k_y}(x,y)\propto \frac{1}{\sqrt{2}}\left (\begin{matrix}\mathrm{sgn}(B) \\ i\end{matrix}\right )\left (e^{-x/\xi_+}-e^{-x/\xi_-}\right )e^{ik_yy}
,\end{equation}
where the penetration depths $\xi_\pm$ now also depend on $k_y$. By projecting the Hamiltonian onto the edge state solution, its energy dispersion is found to depend linearly on momentum
\begin{eqnarray}\label{eq:hedge}
E_{\mathrm{edge}}(k_y)=-\mathrm{sgn}(B)vk_y
,\end{eqnarray}
corresponding to electrons propagating with velocity
\begin{equation}\label{eq:vy}
v_{\mathrm{edge}}=\frac{\partial E_{\mathrm{edge}}(k_y)}{\partial k_y}=-\mathrm{sgn}(B)v.
\end{equation}
Importantly, because $\bd{\psi}_{k_y}(x,y)$ for $B > 0$ ($B < 0$) is an eigenstate of $\sigma_y$ with eigenvalue $+1$ ($-1$), the spin polarisation and the direction of propagation are both determined by $\mathrm{sgn}(B)$. As the spin and the direction of propagation are thus connected, this property is called spin-momentum locking and represents one of the defining features of helical edge states. In particular, electrons with opposite spin counter-propagate.

Before turning to discuss helical edge states in realistic 2d TIs, we want to make the topological distinction between the cases $\mathrm{sgn}(MB)=\pm 1$ more explicit.
Consider the $k$-dependent mass term defined in Eq.~\eqref{eq:diracM(k)}.
At small momenta, the sign of $M(k)$ is determined by $\mathrm{sgn}(M)$, while at large momenta it is determined by $-\mathrm{sgn}(B)$. Then if $\mathrm{sgn}(MB)=-1$ the sign of $M(k)$ at small and large momenta is the same, while in the case $\mathrm{sgn}(MB)=+1$ it changes sign as sweeping through the Brillouin zone (BZ).
The latter (topologically nontrivial) case resembles what we discussed previously, an edge states emerging in the presence mass term which changes sign.

These behaviours do not depend on the specific values of $M$ and $B$, but only on the quantity $\mathrm{sgn}(MB)$ being either $+1$ or $-1$. By considering the area on the Bloch sphere swept by a (normalized) vector $\bd{d}(\bd{k})$ when $\bd{k}$ varies over the entire BZ, a topological invariant can be defined: the Chern number.
To understand this concept, we start from the Berry phase $\gamma_{c,n}$. It is the phase acquired by the Bloch wave function $|u_{n}(\bd{k})\rangle$ (with band index $n$) when the momentum $\bd{k}$ is varied around a closed loop in the BZ. The Berry phase can be expressed either as a line integral of the Berry connection $\bd{\mathcal{A}}_{n}(\bd{k})=i\langle u_n(\bd{k})|\bd{\nabla}_{\bd{k}}|u_{n}(\bd{k})\rangle$ or using Stokes' theorem as the surface integral over the Berry curvature $\bd{\Omega}_n(\bd{k})=\bd{\nabla}_{\bd{k}}\times\bd{\mathcal{A}}_n(\bd{k})$.
The Chern number $n_c$ is defined as the Berry curvature over the BZ, summed over all the occupied bands
\begin{equation}\label{eq:chernn}
n_c=\sum_{n \text{ occ.}}\int_{\mathrm{BZ}} \ d\bd{k} \cdot \bd{\Omega}_n(\bd{k})
,\end{equation}
and it can be shown to be an integer number.
Classes of systems characterized by different Chern numbers cannot be continuously deformed one into another without closing the energy gap, and therefore the gap is ``topologically protected'' against smooth perturbations.
In terms of Eq.~\eqref{eq:chernn}, an analogy can be drawn between the topological classification introduced above and the one characterising different surfaces in terms of their genus $g$. Indeed, the genus is defined as a surface integral of a Gaussian curvature, just like the Chern number is defined as the surface integral of the Berry curvature. Then, the distinction between topologically trivial ($n_c=0$) and non-trivial ($n_c\neq 0$) phases of matter is identical to the different topologies of the sphere with genus $g=0$ and the torus with genus $g=1$.

For the modified Dirac equation, the Chern number calculated with Eq. \eqref{eq:chernn} is
\begin{equation}\label{eq:nc}
n_c=-\frac{1}{2}\left [\mathrm{sgn}(M)+\mathrm{sgn}(B)\right ]
.\end{equation}
Indeed, in the trivial regime, $\mathrm{sgn}(MB) = -1$, the Chern number $n_c=0$, while $n_c=\pm 1$ in the topological regime, where $\mathrm{sgn}(MB)=+1$.

\subsection{BHZ model}\label{sec_1.4}
\begin{figure}[!tt]
\centering
\includegraphics[scale=0.5]{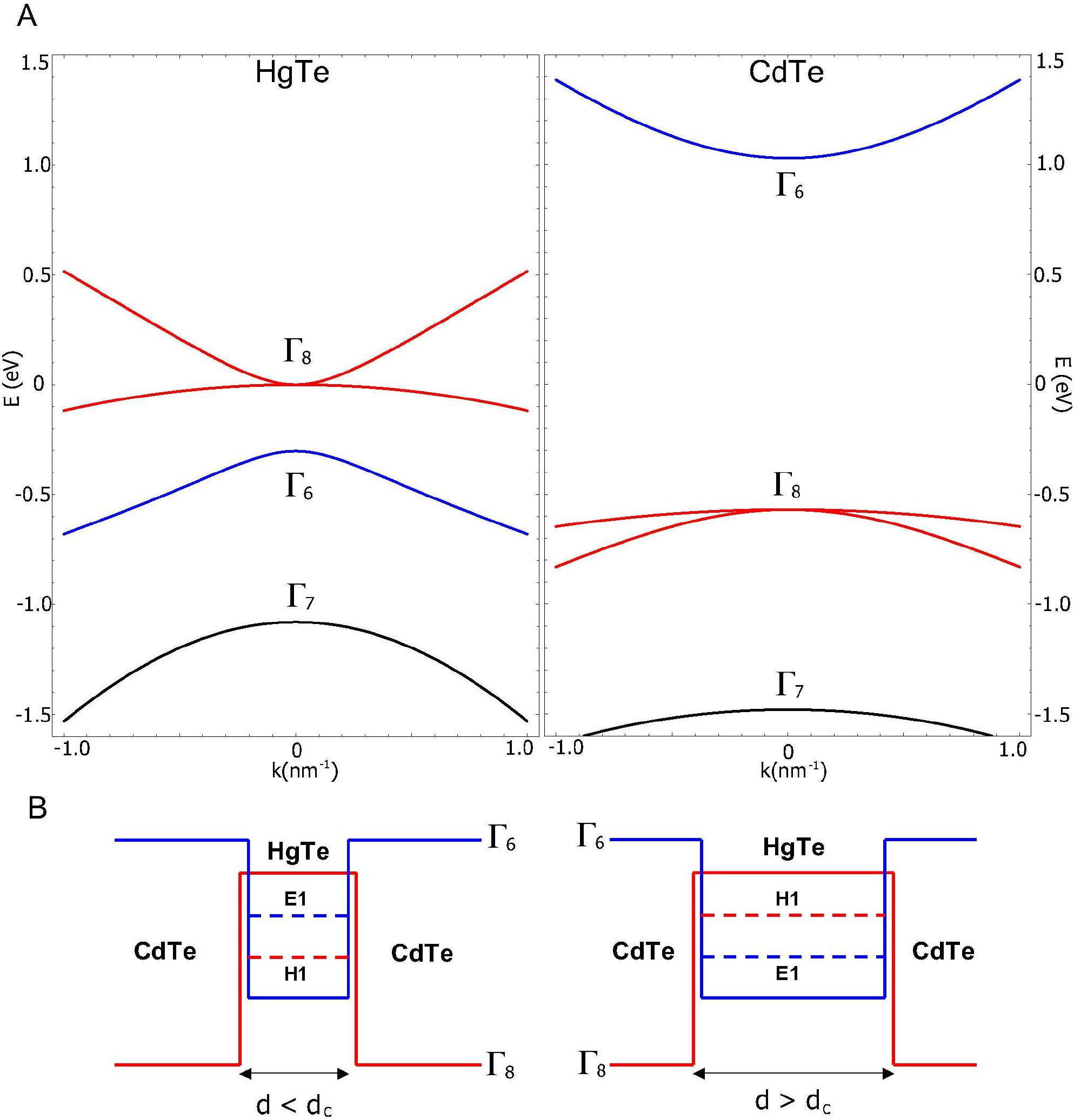}
\caption{(a) Band structure of HgTe and CdTe near the $\Gamma$ point. Note the different band ordering between the $\Gamma_6$ and $\Gamma_8$ bands. (b) The bands of thin QWs (thickness $d<d_c$) resemble the normal ordering of CdTe bulk spectrum, with $M=\left (E_1-H_1\right )/2>0$, while for thick QWs ($d>d_c$) the band ordering is inverted and $M<0$. From Ref.~\cite{qi2011topological} with the courtesy of the authors.}\label{fig:QW}
\end{figure}

The concepts discussed in the previous section are general and can be applied to describe several systems with topological properties. Here we explicitly focus on the first experimentally discovered 2d TI: the HgTe/CdTe QW.
The ideas behind its discovery can be understood as follows.
CdTe has a normal semiconductor band structure, with the valence band $\Gamma_8$, originating from $p$-type bands, separated by an energy gap from the conduction band $\Gamma_6$, which originates from an $s$-type band. On the other hand, because of the strong SOC, bulk HgTe displays an inverted semimetallic band structure, with the $\Gamma_8$ bands above the $\Gamma_6$ valence band, as schematically shown in Fig.~\ref{fig:QW}(a).

Starting from the bulk bands, one can consider different behaviours of thin and thick QWs, whose band structures are determined by confinement, as schemetically shown in Fig.~\ref{fig:QW}(b). When the central layer of HgTe is thin, the energy bands align in a normal ordering, similar to the ones of CdTe. On the other hand, when the width of HgTe is above a critical thickness $d_c$, the energy bands of the QW are expected to be in the inverted regime, similarly to bulk HgTe.
The QW states derived from the heavy-hole $\Gamma_8$ band are denoted by $H_n$,
with $n=1,2,3,\dots$ describing states with an increasing number of nodes.
Analogously, $E_n$ denotes the states derived from the electron $\Gamma_6$ band.
The sign of the energy gap $2M=E_1-H_1$ measured at the $\Gamma$ point between the first valence and conduction bands discriminates between normal ($M>0$) and inverted ($M<0$) regime. The transition between normal and inverted regime occurs at the critical thickness $d_c\approx 6.3$ nm~\cite{bernevig2006quantum}; therefore the experimental fabrication is crucial in order to determine the properties of the QW.

By following Ref.~\cite{bernevig2006quantum}, one can introduce an effective band model, compatible with the symmetries of the system, to describe the QW near the $\Gamma$ point.
We assume the QW to be built along the $\hat{z}$ direction. In the presence of TRS, Kramers theorem states that each state must be doubly degenerate. Then we order the four relevant subbands as $\{|E_1+\rangle, |H_1+\rangle, |E_1-\rangle, |H_1-\rangle\}$, where $|E\pm\rangle$ are Kramers partners, as well as $|H\pm\rangle$, and use them to build an effective low-energy Hamiltonian.
Because of Kramers theorem, terms connecting Kramers partners must vanish.
Furthermore, since $|E_1\pm\rangle$ and $|H_1\pm\rangle$ originate from $s$-like and $p$-like bands respectively, they have opposite parity. Thus, every matrix element connecting these states must be odd under parity. Since we are expanding around the $\Gamma$ point, the most relevant terms must be linear in momentum $k$. For the same reason, diagonal terms can only contain even powers of momentum $k$.
Finally, non-vanishing terms coupling $|E_1\pm\rangle$ and $|H_1\mp\rangle$ states are not admitted, since they would split the Kramers degeneracy via second order perturbation processes. With these considerations, the effective Bernevig-Hughes-Zhang (BHZ) Hamiltonian reads
\begin{equation}\label{eq:TIH}
\bd{H}(\bd{k})=\left (\begin{matrix}\bd{h}(\bd{k}) & 0 \\ 0 & \bd{h}^*(-\bd{k})\end{matrix}\right )
\end{equation}
with the $2\times 2$ matrix
\begin{equation}\label{eq:TIh(h)}
\bd{h}(\bd{k})=\epsilon(k)\bd{\mathbbm{1}}+\bd{d}(\bd{k})\cdot\bd{\sigma}
.\end{equation}
Here, $\epsilon(k)=C-D\left (k_x^2+k_y^2\right )$, $\bd{d}(\bd{k})=\left (Ak_x, -Ak_y, M(k)\right )$, $M(k)=M-B\left (k_x^2+k_y^2\right )$ and $A,B,C,D,M$ are material parameters that depend on the geometrical structure.
The parameter $M$ is indeed half of the energy gap between $|E_1\pm\rangle$ and $|H_1\pm\rangle$ states at the $\Gamma$ point, which depends on the thickness of the QW, as we discussed.

\begin{figure}[!tt]
\centering
\includegraphics[scale=0.3]{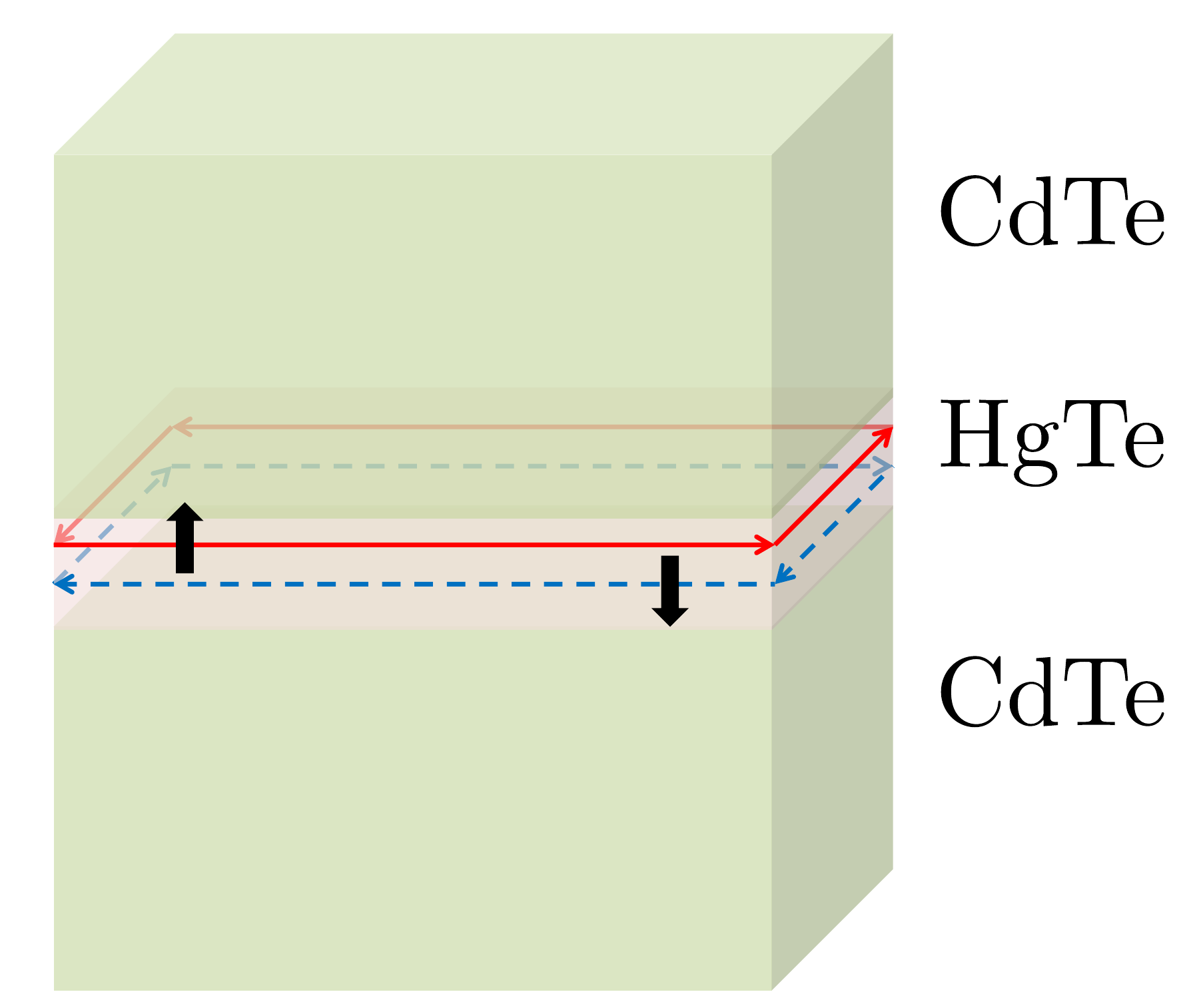}
\caption{In the topological regime, helical edge states exist: electrons with opposite spin-polarisation counter-propagate.}\label{fig:qsh}
\end{figure}

\begin{figure}[!tt]
\centering
\includegraphics[scale=0.5]{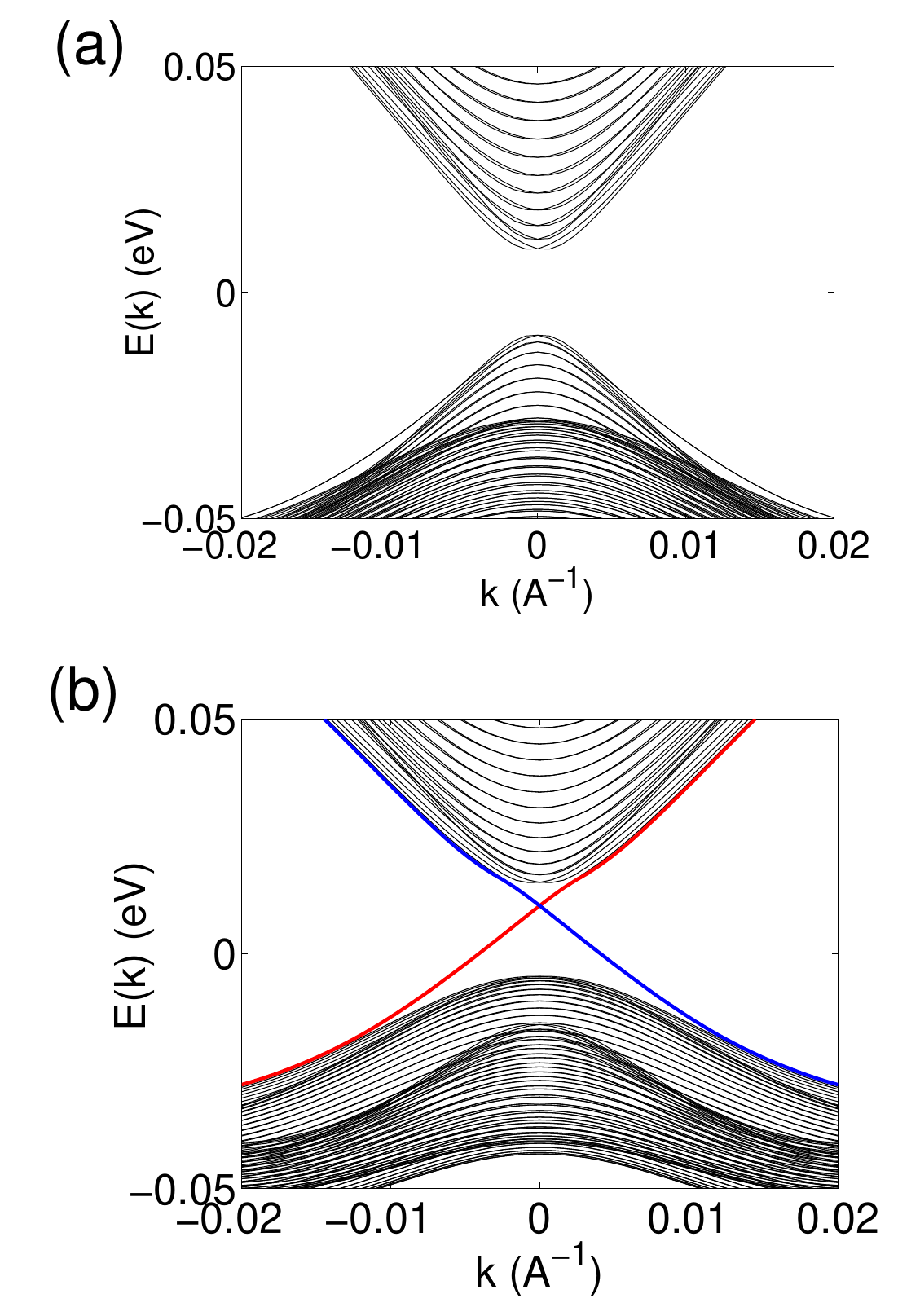}
\caption{Energy spectrum of the HgTe/CdTe QW. (a) In the normal regime $d<d_c$ the system is insulating, with no edge states; (b) in the inverted regime $d>d_c$ the system behaves as a topological insulator, with insulating bulk states and gapless, linear dispersing helical edge states. From Ref.~\cite{qi2011topological} with the courtesy of the authors.}\label{fig:bands}
\end{figure}

\noindent If we neglect $\epsilon(k)$, a comparison with Eqs.~\eqref{eq:Hchern}-\eqref{eq:diracM(k)} indeed reveals a modified Dirac Hamiltonian with $v=A$.
Just like the modified Dirac Hamiltonian, the individual blocks \eqref{eq:TIh(h)} both break TRS. However, TRS is restored in the four band model \eqref{eq:TIH}.
Indeed, the QW can be described by two decoupled modified Dirac Hamiltonians $\bd{h}(\bd{k})$ and $\bd{h}^*(-\bd{k})$ related by TRS.
Therefore, two states appear at the sample edge. They are Kramers partners, have opposite spins and counter-propagate: the helical edge states, schematically shown in Fig.~(\ref{fig:qsh}), characterised by spin-momentum locking.
The Chern numbers $n_-$ and $n_+$ associated to $\bd{h}(\bd{k})$ and $\bd{h}^*(-\bd{k})$ respectively are
$n_{\pm}=\pm[\mathrm{sgn}(M)+\mathrm{sgn}(B)]/2$. Despite the total Chern number $n_c=n_++n_-$ being zero, a nontrivial spin Chern number can be defined
\begin{equation}\label{eq:chernspin}
n_s=\frac{n_+-n_-}{2}
\end{equation}
which vanishes in the topologically trivial case, $\mathrm{sgn}(MB)=-1$, but takes on the value $n_s = \pm 1$ in the topological one, $\mathrm{sgn}(MB)=+1$.
This reflects again the $\mathbb{Z}_2$ topological order predicted by Kane and Mele~\cite{kane2005z}.
Note that since the parameter $B<0$ in HgTe/CdTe QWs, the topological phase corresponds to the inverted regime $M<0$ of thick ($d>d_c$) QWs, while the normal regime $M>0$ corresponding to thin QWs ($d<d_c$) is trivial.
We will see in Sec.~\ref{sec:exp} how these theoretical predictions can be tested experimentally.
Analogously to Eq.~\eqref{eq:hedge}, the helical edge states have a Dirac spectrum,
\begin{equation}\label{eq:hedgek}
E_{\uparrow,\downarrow}(k_y)=\pm v_Fk_y,
\end{equation}
where $v_F = A$, and for HgTe/CdTe QWs, $v_F\sim 5\cdot 10^5$ m$/$s.
The QW band structure in the normal and inverted regime are plotted in Fig.~\ref{fig:bands}.
A bulk energy gap $\Delta\approx 30$ meV exists in both cases, separating the bulk valence and conduction bands. However, a pair of gapless dispersing states appear in the inverted regime, corresponding to the energy spectrum Eq. \eqref{eq:hedgek} of two counter-propagating edge modes.

\begin{figure}[!tt]
\centering
\includegraphics[scale=0.5]{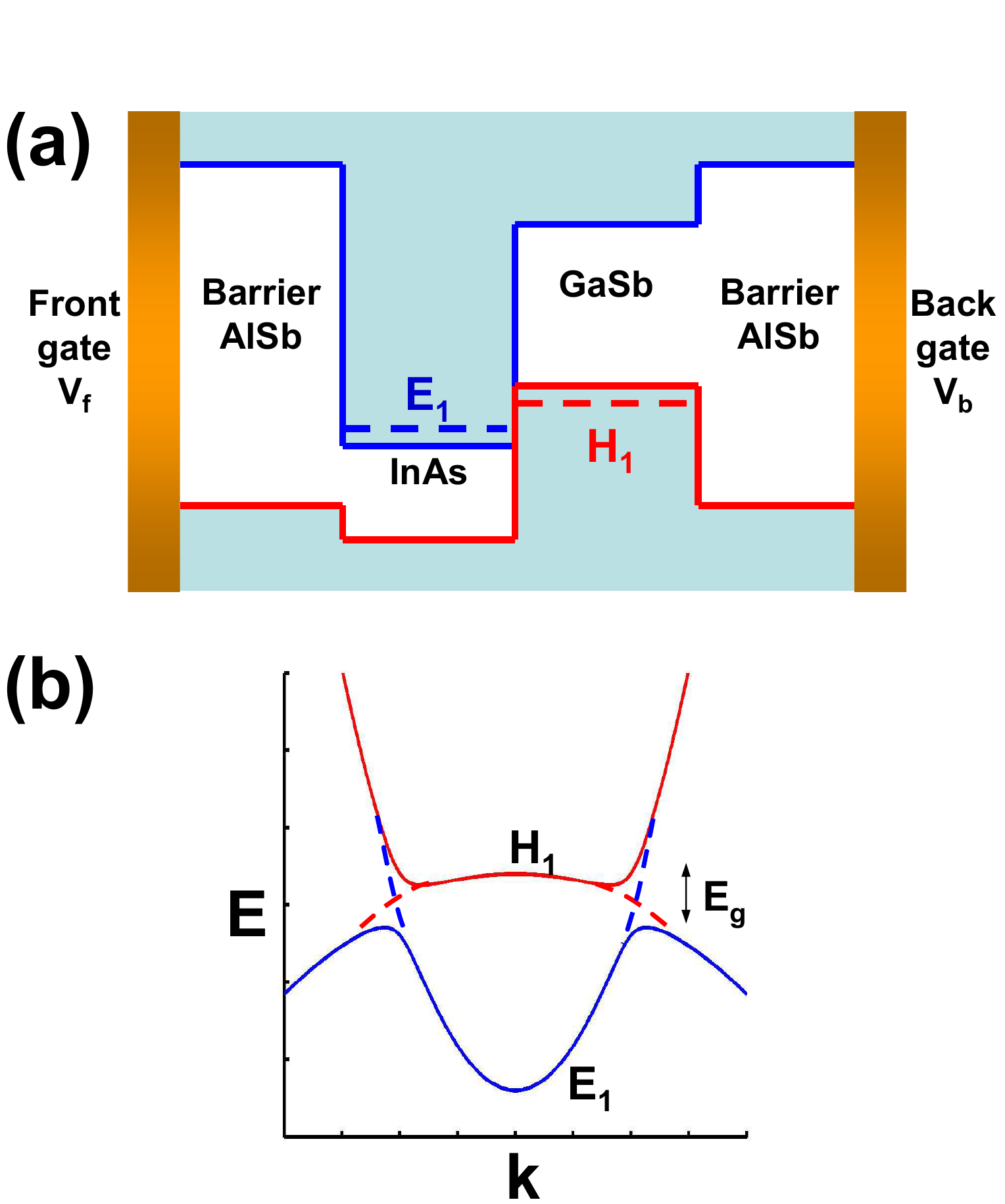}
\caption{(a) Quantum well layer structure: the $E_1$ valence subband is localized in the InAs layer, with the conduction subband $H_1$ in the GaSb layer. (b) When electron and hole subbands cross each other the inverted regime is achieved: because of hybridization at the crossings, a bulk gap opens ($E_{\mathrm{g}}$) which protects the existance of edge states. From Ref.~\cite{Liu2008quantum} with the courtesy of the authors.}\label{fig:Liu08_QW}
\end{figure}

A possible disadvantage of HgTe/CdTe QWs is that the topological phase transition, depending essentially on the thickness, cannot be changed \emph{in situ}.
This limitation pushed theorists to search for more versatile heterostructures displaying the QSH effect.
Liu \textit{et al.} theoretically demonstrated the occurrence of QSH effect in type-II inverted semiconductors~\cite{Liu2008quantum}.
The proposed heterostructure consist of asymmetric layers of AlSb/InAs/GaSb/AlSb as shown in Fig.~\ref{fig:Liu08_QW}(a).

\begin{figure}[!tt]
\centering
\includegraphics[scale=0.75]{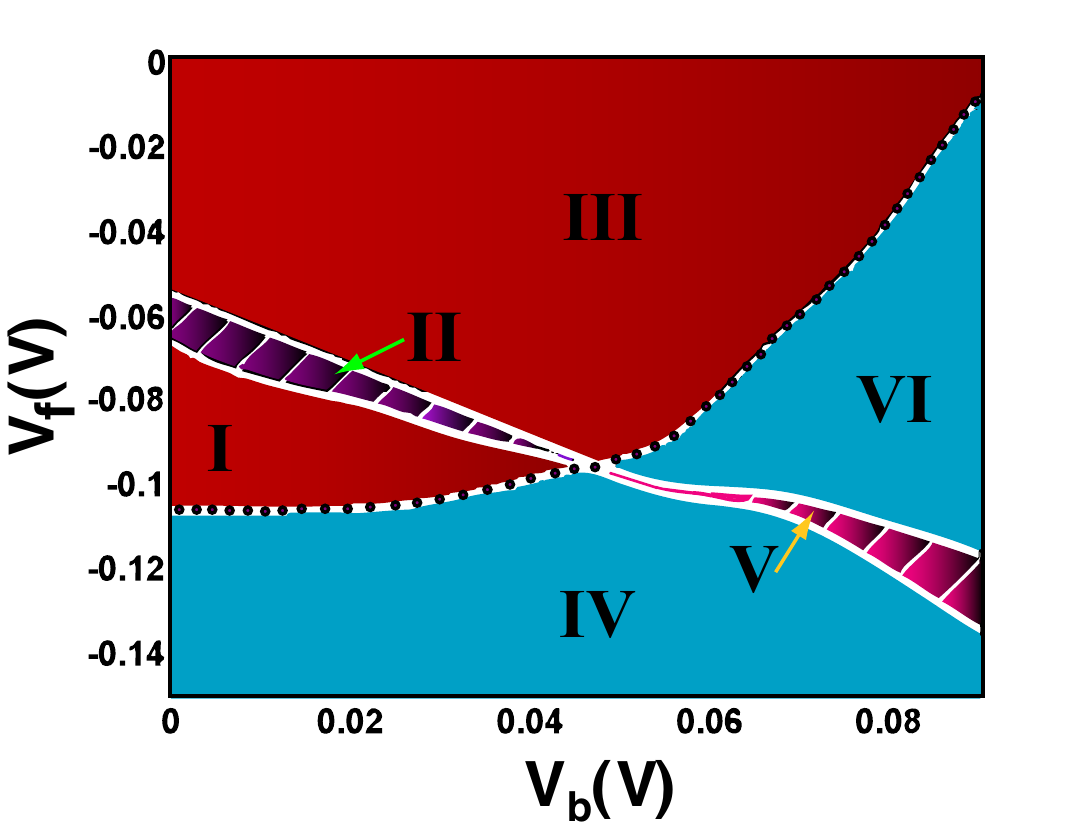}
\caption{Phase diagram for a 30-10-10-30 $\mu$m thick QW in Fig.~\ref{fig:Liu08_QW}(a) as a function of front ($V_{\mathrm{f}}$) and back ($V_{\mathrm{b}}$) gate voltages, defined with respect to the Fermi level of the QW. Regions I, II, III (IV, V, VI) correspond to the inverted (normal) regime. Regions I and IV (III and VI) are in the $p(n)$-doped regime, while the Fermi energy is tuned within the inverted (normal) bulk energy gap in region II (V). Therefore the QSH phase appears in region II. From Ref.~\cite{Liu2008quantum} with the courtesy of the authors.}\label{fig:Liu08_gate}
\end{figure}

Crucially for the topological properties, the valence band of GaSb is higher in energy than the conduction band of InAs: therefore the conduction and valence bands are spatially separated, the former being in the InAs layer, the latter in the GaSb layer. Following the BHZ model, one can focus on the nearly degenerate lowest electron and hole subbands $H_1$ and $E_1$, with the other bands well separated in energy.
By increasing the layer thickness, a level crossing between $E_1$ and $H_1$ occurs, putting the system into the inverted regime. Hybridization between electron and hole-like subbands leads to the opening of a gap $E_{\mathrm{g}}$, making the bulk of the system insulating, as shown in Fig.~\ref{fig:Liu08_QW}(b). By following the ideas of the BHZ model, the QW should then behave as a trivial insulator in the normal regime, while displaying QSH physics in the inverted one.
A major advantage of such a heterostructure compared to HgTe/CdTe QWs is that back gate and front gate voltages, placed as in Fig.~\ref{fig:Liu08_QW}(a), can be used to modify the band alignment and simultaneously adjust the Fermi energy, as shown in Fig.~\ref{fig:Liu08_QW}(a).
Therefore a rich phase diagram emerges as a function of gate voltages, see Fig.~\ref{fig:Liu08_gate}.

In particular, by tuning the gate voltages in region II in Fig.~\ref{fig:Liu08_gate} the system reaches the QSH phase with the Fermi energy tuned inside the inverted bulk energy gap, where the presence of protected helical edge states can be investigated. Furthermore, it is possible to electrostatically tune the system either in the inverted (regions I, II, III) or in the normal (regions IV, V, VI) regime, allowing to investigate the nature of the topological phase transition.

\begin{figure}[!tt]
\centering
\includegraphics[scale=0.3]{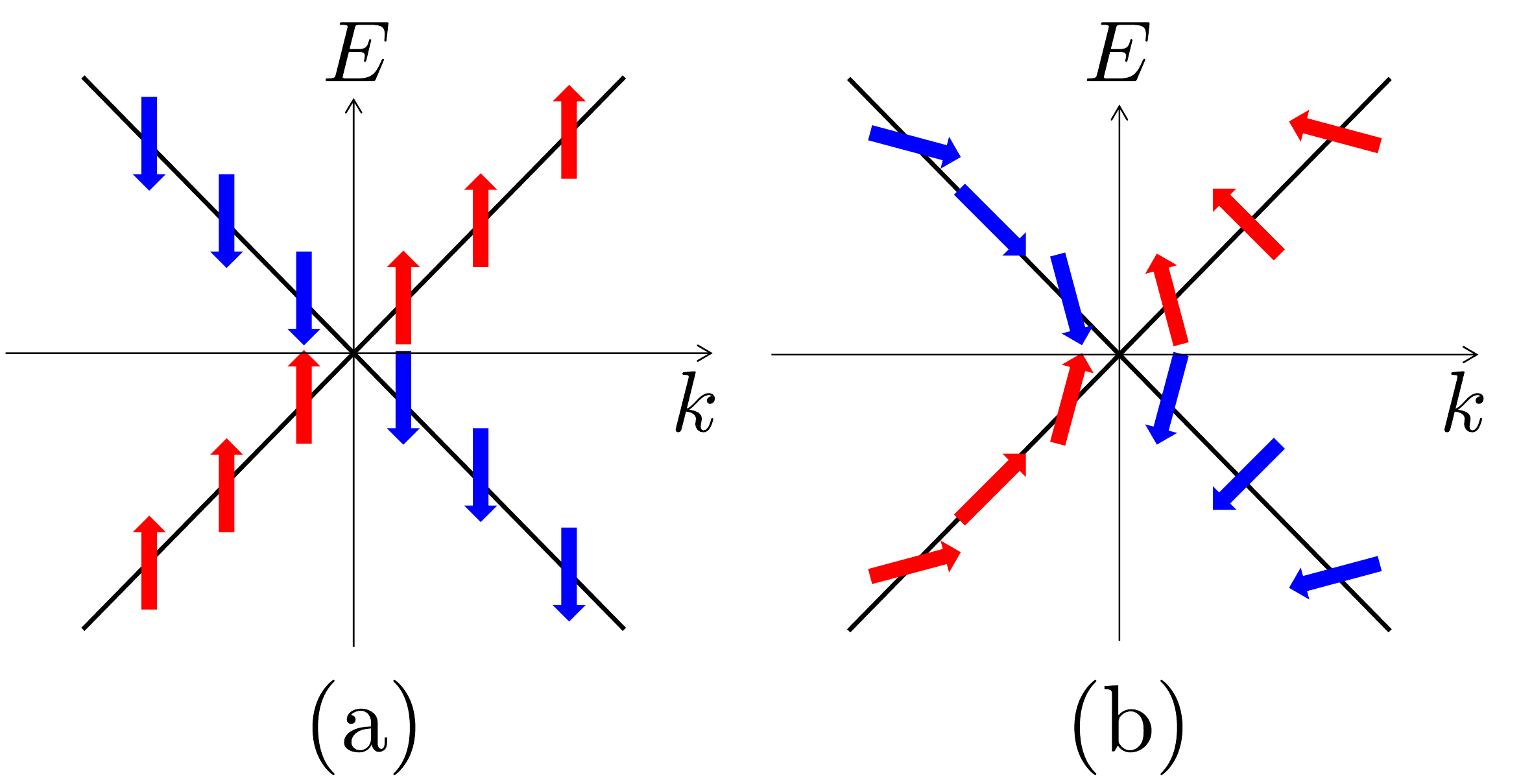}
\caption{(a) The edge states predicted by the BHZ model~\eqref{eq:TIH} have well-defined spin polarisation. (b) In the presence of BIA or SIA, the spin polarisation is no longer constant, but at any energy the Kramers partners still have opposite momenta and spins.}\label{fig:generic}
\end{figure}

The BHZ model predicts the existence of 1d channels at the boundaries of a 2d TI, where electrons with opposite spin counter-propagate. This result has been achieved starting from the simplified Hamiltonian Eq. \eqref{eq:TIH}, where bulk inversion asymmetry (BIA) and structural inversion asymmetry (SIA) have been neglected~\cite{zhang2009topological}.
This assumption is not always justified.
For example, in InAs/GaAs heterostructures both SIA and BIA are always present, and also in HgTe/CdTe QW the axial symmetry may be lifted, for example, by the application of an electric field in the $\hat{z}$ direction~\cite{zhang2009topological, rothe2010fingerprint}, while in lattice models like silicene the axial spin symmetry is broken by Rashba spin-orbit coupling terms.
Therefore, a more general Hamiltonian takes into account SIA and BIA terms,
\begin{eqnarray}
\bd{H}_{\mathrm{BIA}}(\bd{k})&=&\left (\begin{matrix}
0 & 0 & \Delta_e(k_x+ik_y) & -\Delta_0 \\
0 & 0 & \Delta_0 & \Delta_h(k_x-ik_y) \\
\Delta_e(k_x-ik_y) & \Delta_0 & 0 & 0 \\
\Delta_0 & \Delta_h(k_x+ik_y) & 0 & 0
\end{matrix} \right),\\
\bd{H}_{\mathrm{SIA}}(\bd{k})&=&\left (\begin{matrix}
0 & 0 & i\xi_e(k_x-ik_y) & 0 \\
0 & 0 & 0 & 0 \\
-i\xi_e^*(k_x+ik_y) & 0 & 0 & 0 \\
0 & 0 & 0 & 0
\end{matrix} \right),
\end{eqnarray}
where we kept terms up to linear order in momenta.
Of course, the inclusion of such terms in the Hamiltonian does not change the main conclusions we had drawn about the existence of a topological phase transition.
Indeed, its existence is protected by TRS, which is not compromised by either $\bd{H}_{\mathrm{BIA}}$ or $\bd{H}_{\mathrm{SIA}}$.
However, since both BIA and SIA couple spin-up and spin-down blocks of the Hamiltonian, spin is no longer conserved, so the counter-propagating edge modes do not have a well defined spin polarisation any more~\cite{rothe2010fingerprint, maciejko2010magnetoconductance}. In general, the correct picture is no longer given by Fig.~\ref{fig:generic}(a), with constant spin polarisation on a given dispersion line, but rather resembles Fig.~\ref{fig:generic}(b)~\cite{schmidt2012inelastic, orth2013point, Kainaris2014conductivity, rod2015spin}.

At a fixed energy, the two states forming a Kramers doublet still have opposite momenta and opposite spins, but the spin polarisation, e.g., the spinor in wave functions as Eq.~\eqref{eq:boundstate2}, changes with momentum.
Therefore, the more general notion of generic helical liquids was introduced~\cite{schmidt2012inelastic, rod2015spin} as a label for a helical edge state without axial spin symmetry. The distinction between systems with conserved and non-conserved spin is usually not important if only elastic scattering processes are relevant. For instance, the zero-temperature conductance quantization is guaranteed by TRS. However, a spin axis rotations leads to a nonzero overlap between states with momenta $k$ and $-k'$ for $k' \neq k$, and this has an impact on inelastic scattering processes. It thus affects, for instance, the conductance of an edge state at finite temperatures. This will be discussed in Sec.~\ref{sec:scattering}.

\subsection{Topological protection of the helical edge states}\label{subsec:protection}
If the Fermi energy is tuned into the bulk energy gap of a trivial insulator electron transport is inhibited. On the other hand, when tuning the Fermi energy into the bulk energy gap of a topological insulator, electron transport remains possible via the edge states.
In general, the presence of impurities or defects in one-dimensional channels is very detrimental for the conductance, suppressing the elastic mean free path because of the possibility of electron backscattering.

\begin{figure}[!tt]
\centering
\includegraphics[scale=0.3]{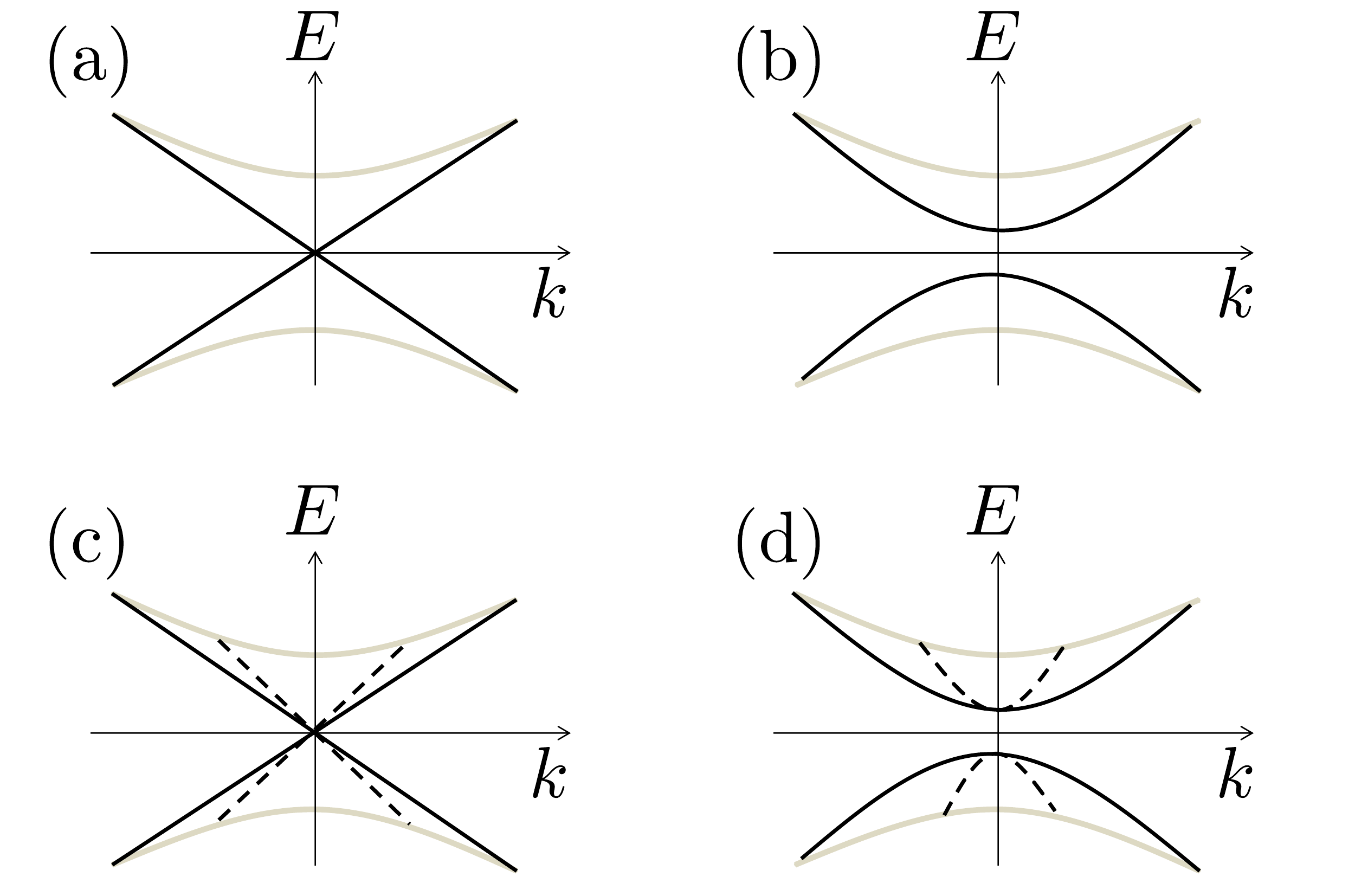}
\caption{If an odd number of Kramers doublets is present on the edge (a), a gap in the edge state spectrum cannot be opened without violating the Kramers theorem (b); on the other hand, if an even number of Kramers doublets is present (c), the edge states can be fully gapped out also in the presence of TRS (d).}\label{fig:Kramers}
\end{figure}

The 1d edge states of topological insulators, in contrast, are topologically protected as long as TRS is preserved. Therefore, ballistic transport, insensitive to impurities and disorder, is predicted.
This argument can be made more rigorous by recalling the Kramers theorem: in the presence of TRS, the eigenstates come in Kramers doublets. Under TRS $\bd{k}\to -\bd{k}$ and $\bd{\sigma}\to -\bd{\sigma}$, implying that at the time-reversal invariant $\Gamma$ point, the Kramers partners must be degenerate.
The topological protection depends on the number of Kramers pairs appearing at the edge and reflects the $\mathbb{Z}_2$ topological order of the QSH effect: if an odd number of Kramers doublets is present, it is impossible to gap out all of them without violating the Kramers theorem, as shown in Fig.~\ref{fig:Kramers}(a-b); on the other hand, if an even number of Kramers doublets is present, one can gap out all of them without violating the Kramers theorem, as shown in Fig.~\ref{fig:Kramers}(c-d).

If it is possible to gap out the all the Kramers partners without breaking TRS, then the state would be topologically equivalent to a trivial insulator $n_s=0$. In the same way, a topological insulator with $n_s=\pm 1$, corresponding to an odd number of Kramers partners, cannot be adiabatically connected to a trivial insulator $n_s=0$ as long as TRS is preserved.

The Kramers theorem ensures the protection of the helical edge states from elastic and non-magnetic backscattering, but does not prevent the system to become localised by other mechanisms, as will be discussed later in Sec.~\ref{sec:scattering}. In particular, it turns out the interactions can open a gap in the spectrum without explicit breaking of TRS.

\section{Experimental evidences of the helical edge states}\label{sec:exp}
A possible way to identify ballistic electron transport along the sample edges is the quantized resistance, which contrasts with the non-universal resistance Ohm's law would predict for the diffusive 2d bulk transport. If the Fermi energy lies within the bulk energy gap and the temperature is low enough, only edge states will be able to contribute to transport, and the Landauer-Buttiker formalism can be used to predict the multi-terminal resistance.
The relationship between the current flowing out of the $i$th contact due to voltages $\{V_j\}$ applied at the contacts is
\begin{equation}\label{eq:LandButt}
I_i=\frac{e^2}{h}\sum_{j}\left (T_{ji}V_i-T_{ij}V_j\right )
,\end{equation}
with $T_{ji}$ the transmission probability from the $i$-th to the $j$-th contact.
In the presence of TRS the transmission matrix is symmetric $T_{ji}=T_{ij}$.
Moreover $T_{i,i+1}=T_{i+1,i}=1$ because backscattering between Kramers partners is inhibited, while the other entries of the transmission matrix are zero.

\begin{figure}[!tt]
\centering
\includegraphics[scale=1]{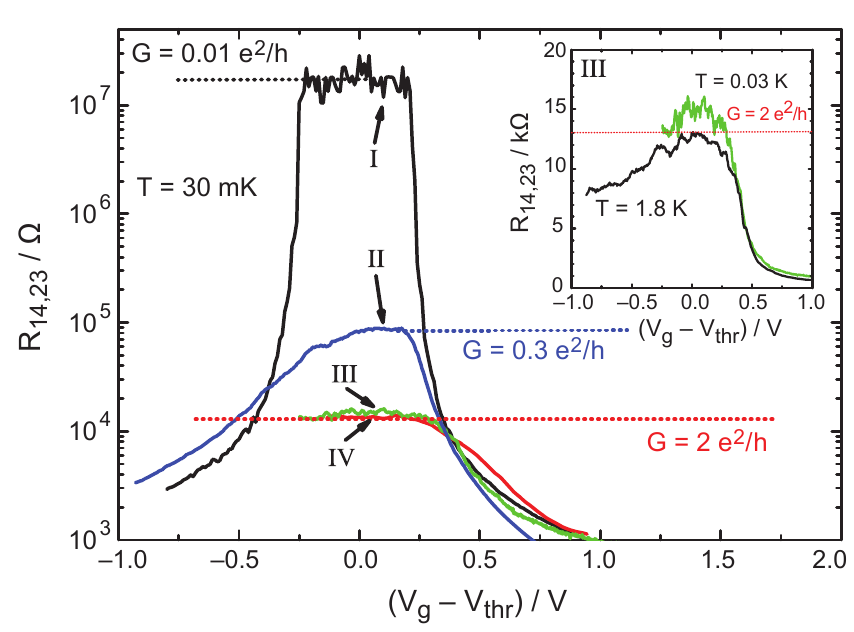}
\caption{Four-terminal resistance $R_{14,23}$ as a function of gate voltage $V_{\mathrm{g}}$ that allows to change the Fermi energy; for $-0.5$ V$\apprle V_{\mathrm{g}}-V_{\mathrm{thr}}\apprle 0.5$ V the Fermi energy lies in the bulk energy gap. The measurements are performed in the absence of external magnetic field at temperature $T=30$ mK. Different curves correspond to different samples: (I, black) device size $(20.0 \times 13.3)$ $\mu$m$^2$ in the normal regime ($d<d_c$); (II, blue) device size $(20.0 \times 13.3)$ $\mu$m$^2$ in the inverted regime ($d>d_c$); (III, green) device size $(1.0 \times 1.0)$ $\mu$m$^2$ in the inverted regime ($d>d_c$); (IV, red) device size $(1.0 \times 0.5)$ $\mu$m$^2$ in the inverted regime ($d>d_c$). Inset: Four-terminal resistance $R_{14,23}$ as a function of gate voltage for two different temperatures $T=30$ mK (green) and $T=1.8$ K (black). From Ref.~\cite{konig2007quantum} with the courtesy of the authors.}
\label{fig:konig_quant}
\end{figure}

The first experimental evidence of transport occurring on the edge of a 2d TI was provided in 2007 \cite{konig2007quantum}.
In the presence of protected edge states, Eq.~(\ref{eq:LandButt}) predicts a four-terminal resistance $R_{14,23}=h/2e^2$, which is in good agreement with the experimental results shown in Fig.~\ref{fig:konig_quant} (red and green curves). When the Fermi energy is inside the bulk energy gap, the resistance is very close to the quantum plateau.
Note that the same plateau is realised by QWs with different sample sizes, provided that $d>d_c$. This suggests that the transport indeed occurs via edge states and not through the bulk.
However, if the distance between the contacts is greater than the inelastic mean free path, estimated to be few $\mu$m, some sort of scattering mechanism takes place and the form of the scattering matrix becomes non-universal, leading to deviations from the predicted plateau. This mechanism explains the different behaviour of the blue curve in Fig.~\ref{fig:konig_quant}, corresponding to an inverted QW whose sample size exceeds the inelastic mean free path. We devote Sec.~\ref{sec:scattering} to the discussion of the possible backscattering processes.
The black curve of Fig.~\ref{fig:konig_quant} shows that edge states disappear in the normal, non-topological regime $d<d_c$, where the QW behaves as a trivial insulator, whose resistance saturates inside the bulk energy gap.
This scenario has been confirmed in other multi-terminal transport experiments~\cite{roth2009nonlocal}.
Finally, the quantum plateau appears to be only weakly sensitive to temperature variations, as shown in the inset.

\begin{figure}[!tt]
\centering
\includegraphics[scale=1]{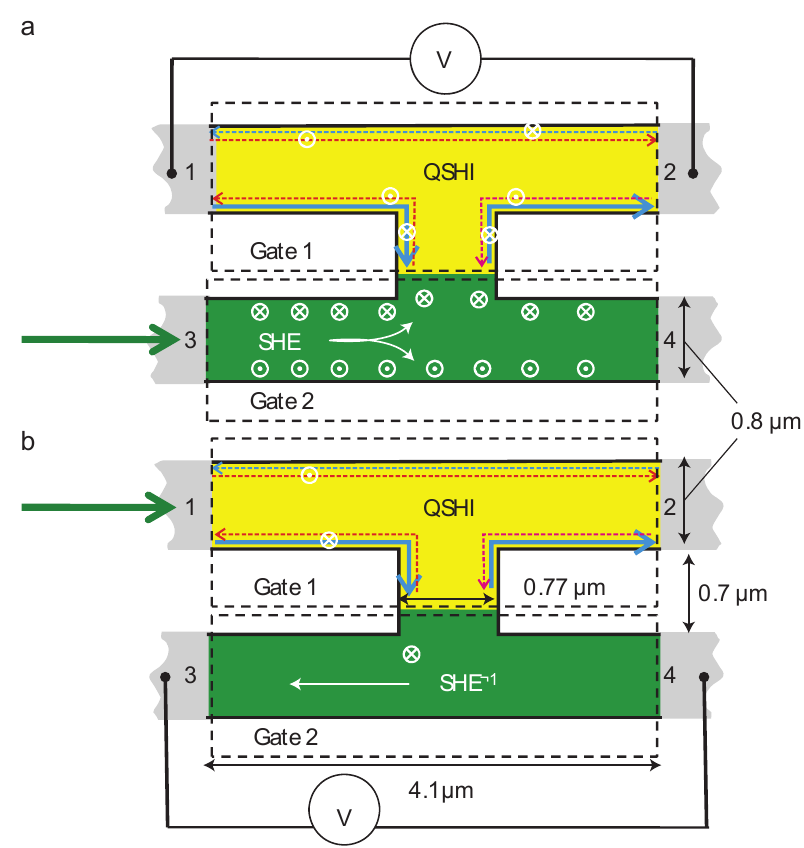}
\caption{Schematic of the experimental setup fabricated by Br\"{u}ne \textit{et al.}~\cite{brune2012spin}, with two T-shaped bars, one of which is in the QSH regime (yellow) and the other one is metallic and shows the spin Hall effect (green), combined to form a hybrid H-shaped bar. (a) Current is injected in the metallic region through contacts $3$ and $4$, and, because of spin Hall effect, spin imbalance accumulates at the edges of the leg; here the spin imbalance is transferred, via the spin-polarised helical edge states of the QSH T-bar, to contacts $1$ and $2$, where a finite potential difference is measured. (b) Injectors and detectors are interchanged: current is injected in the topological region through contacts $1$ and $2$, carried by the spin-polarised edge states, causing spin accumulation in the lower part of the central leg, which can be detected via the inverse spin Hall effect as a voltage difference which develops between contacts $3$ and $4$. From Ref.~\cite{brune2012spin} with the courtesy of the authors.}
\label{fig:brune}
\end{figure}

The measurements of non-local transport properties confirm the physical picture of two counter-propagating channels where backscattering is forbidden, but do not provide evidence of their spin-polarisation. To shed light on this point, experimentalists \cite{brune2012spin} used a split-gate technique to combine two T-shaped bars, one in the QSH regime and the other one in the non-topological spin Hall regime, to fabricate a hybrid H-bar, as shown in Fig.~\ref{fig:brune}.

Two protocols are implemented to test the spin polarisation of the helical edge states. In configuration (a), a current is injected in the metallic region from contact $3$ to $4$, with the contacts $1$ and $2$ used as voltage probes, while the opposite happens in configuration (b).
When a charge current is injected into the metallic region (a), spin-up and spin-down electrons accumulate on different edges because of the spin Hall effect. This imbalance is transferred to the confining QSH region: then, only if the helical edge states are spin-polarised, a finite chemical potential difference between terminals $1$ and $2$ is induced. Analogously, if a current flows between terminals $1$ and $2$ (b), a finite chemical potential difference between contacts $3$ and $4$ is expected only if the edge states are spin-polarised.
In both configurations, Br\"{u}ne and coworkers measured finite and nearly quantised multi-terminal resistances $R_{34,12}$ (a) and $R_{12,34}$ (b), thus providing evidence of the spin polarisation of the helical edge states.

A few years after its discovery in HgTe/CdTe, the QSH effect was also observed in InAs/GaSb heterostructures. Unequivocal signatures of the presence of helical edge states in the inverted regime were found via transport measurements~\cite{knez2011evidence, du2015robust}.

\begin{figure}[!tt]
\centering
\includegraphics[scale=1]{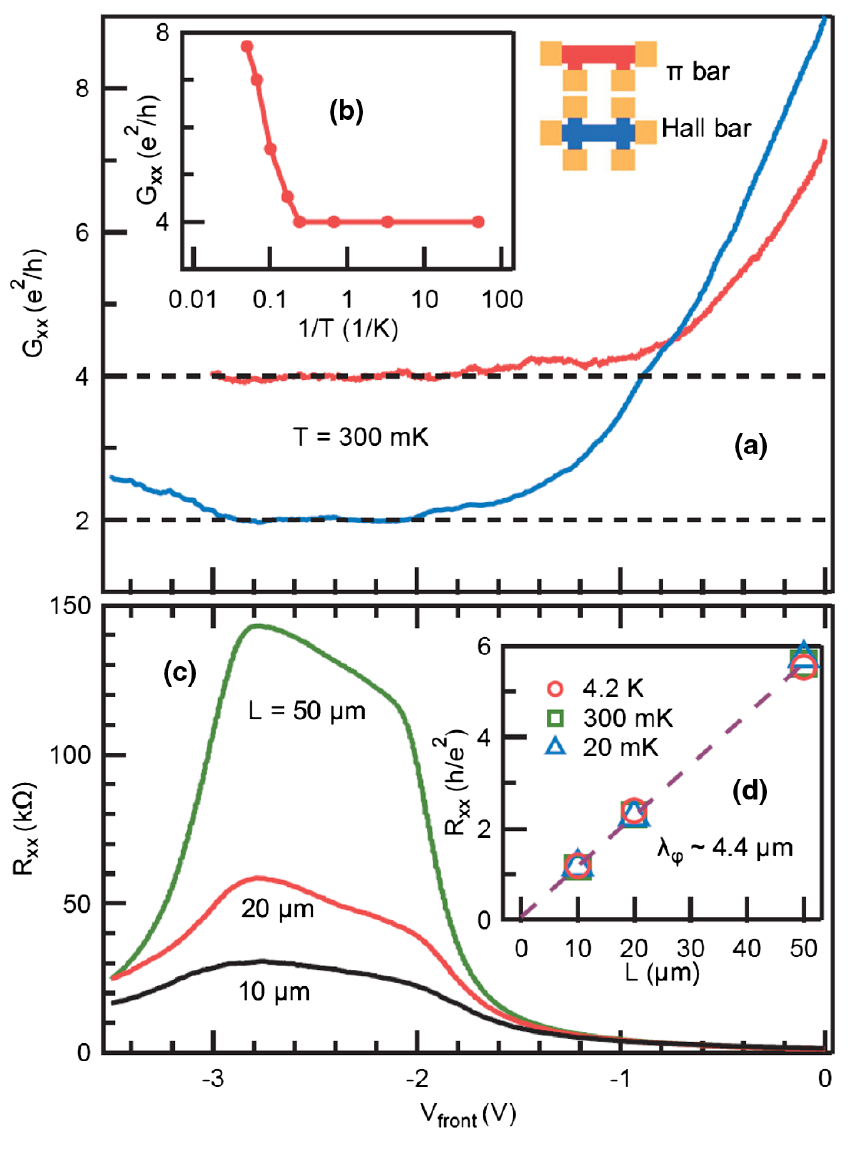}
\caption{(a) Quantized plateaus of $G_{xx}$, $2e^2/h$ and $4e^2/h$, in two different $2$ $\mu$m long and $1$ $\mu$m wide devices, shown on top right, corresponding to $\pi$ (red curve) and Hall (blue curve) bars. (b) The quantized plateaus persist in the range $20$ mK to $4$ K, where conductance starts increasing due to delocalized bulk carriers. (c) Deviations from the quantized plateau emerge when the device length exceeds few $\mu$m, indicating that some mechanism of backscattering take place. (d) Out of the ballistic regime, the edge resistance scales linearly with the length of the sample, indicating a phase coherent length of $\lambda_{\varphi}\sim 4.4$ $\mu$m. Also in the diffusive regime, the edge transport appears to be temperature independent in a wide range from $20$ mK to $4$ K. From Ref.~\cite{du2015robust} with the courtesy of the authors.}
\label{fig:Du15_T}
\end{figure}
Figure~\ref{fig:Du15_T}(a) shows the quantum plateau in two different setups: in the Hall bar a quantized conductance $2e^2/h$ is observed, while the plateau reaches a value of $4e^2/h$ in the $\pi$-bar. These results are in perfect agreement with the Landauer-B\"{u}ttiker theory~\eqref{eq:LandButt}.
Analogously to what happens in HgTe/CdTe, the onset of some backscattering mechanism is observed in longer samples, see Fig.~\ref{fig:Du15_T}(c). As in HgTe/CdTe devices, a coherence length of few $\mu$m is observed.
The edge states of InAs/GaSb seem to be very weakly temperature dependent: for temperatures lower than $4$ K, where bulk states become delocalized affecting the total conductance, see Fig.~\ref{fig:Du15_T}(b), the edge state conductance appears temperature independent in a wide range, see also Fig.~\ref{fig:Du15_T}(d).

\begin{figure}[!tt]
\centering
\includegraphics[scale=1]{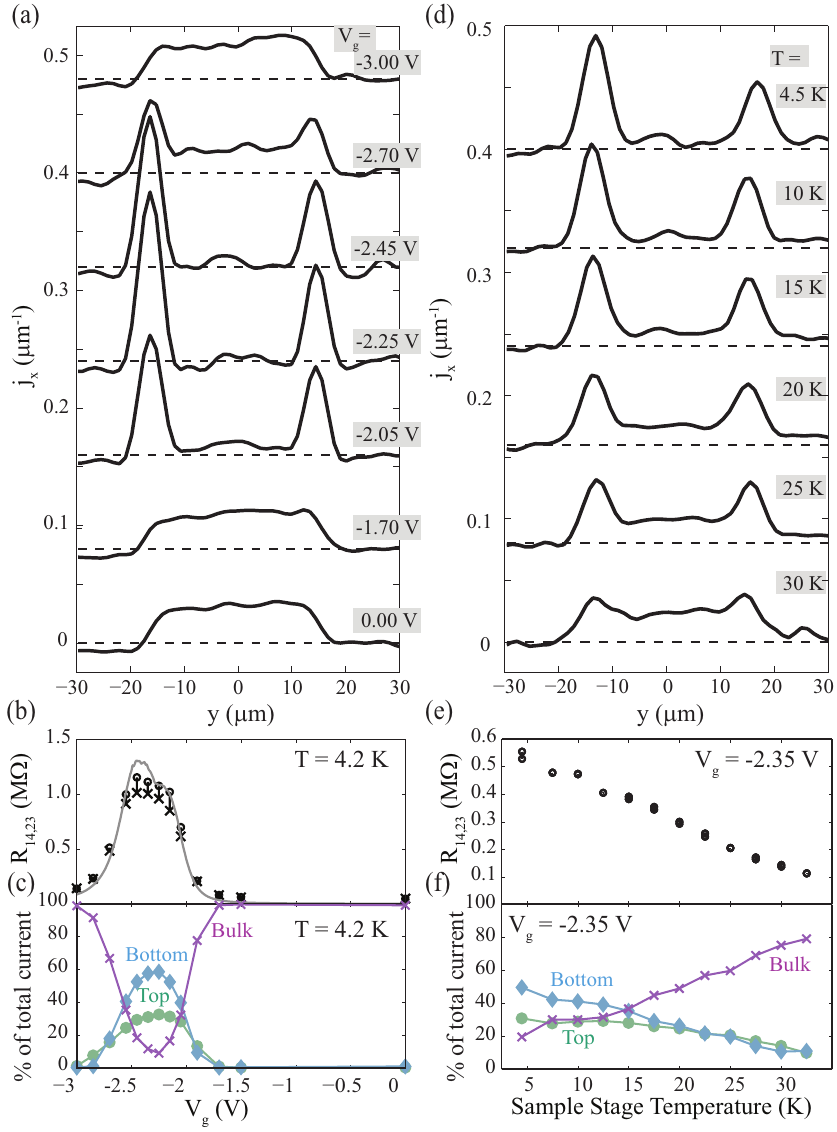}
\caption{Imaging edge and bulk currents in InAs/GaSb heterostructures as a function of gate voltage $V_{\mathrm{g}}$ (a-c) and temperature $T$ (d-e).
(a) By tuning the gate voltage $V_{\mathrm{g}}$, a crossover from bulk dominated to edge dominated transport is observed.
(b) Gate voltage dependence of the four-terminal resistance in a downward gate voltage sweep before imaging the current (gray line) and immediately before (open circles) and after (crosses) each image in a subsequent sweep.
(c) Fitted percentages of current flowing in the two edges and in the bulk as a function of the gate voltage.
(d) The edge states persist in a wide range of temperatures, up to $\approx$ $30$ K, while the bulk contribution becomes more and more relevant by increasing the temperature.
(e) Four-terminal resistance decreases as a function of temperature, mostly due to activated conductive bulk states.
(f) Fitted percentages of current flowing in the two edges and in the bulk as a function of temperature.
In panels (a) and (d) $y=0$ corresponds to the center of the sample, and the zero of each profile is represented by the dashed line.
From Ref.~\cite{spanton2014images} with the courtesy of the authors.}
\label{fig:spanton}
\end{figure}

In addition, the edge states of 2d TIs have been mapped in real space by imaging techniques.
By using a scanning superconducting quantum interference device (SQUID) it has become possible to image edge and bulk contributions to transport in HgTe/CdTe \cite{nowack2013imaging} and InAs/GaSb \cite{spanton2014images} systems.
This technique allows a direct visualization of the position of the dominant transport channels in a 2d TI.
Figure~\ref{fig:spanton}(a-c) shows the evolution from bulk dominated to edge dominated transport in InAs/GaSb systems as a function of gate voltage: when $V_{\mathrm{g}}$ brings the Fermi level into either the valence ($V_{\mathrm{g}}=0$ V) or the conduction band ($V_{\mathrm{g}}=-3$ V), transport mainly occurs in the bulk. Between these two situations, when the Fermi energy lies in the bulk gap, edge transport dominates. It should be pointed out that the widths of the peaks in Fig.~\ref{fig:spanton} reflect the limited measurement resolution and not the actual widths of the edge states.
Again, the edge states appear very stable against temperature, as shown in Fig.~\ref{fig:spanton}(d-f), up to a surprisingly high temperature of about $30$ K.

The picture of gapless helical edge states coexisting with insulating bulk states is further corroborated by experiments involving hybrid superconducting-TI devices. The interest in such systems is mainly owed to the predicted emergence of Majorana bound states, whose manipulation and control could be beneficial for quantum computation~\cite{alicea2012new, beenakker2011search, leijnse2012introduction, Nayak2008non}. Although at the initial stage experiments have been performed by exploiting induced superconductivity in quasi-helical quantum wires~\cite{mourik2012signatures, das2012zero}, recently different experimental groups have successfully investigated superconductivity in 2d TI based platforms.
In 2012 Knez \textit{et al.} observed perfect Andreev reflection in SC-QSHI-SC junctions using InAs/GaSb as a QSHI~\cite{knez2012andreev}, thus providing an alternative signature of the presence of helical edge states~\cite{adroguer2010probing}.
More recently, two different experimental groups have managed to induce superconductivity in 2d TIs, both in HgTe/CdTe~\cite{hart2014induced} and in InAs/GaSb~\cite{pribiag2014edge} systems. The topological nature of the edge supercurrents was further supported by tuning the system in and out the topological regime and showing that the supercurrents vanish when the TIs were in the topologically trivial regime.
Although the ultimate goal of this research is to achieve useful platforms for quantum computation, they already now strongly contribute to shedding light on the nature of topological insulators and their edge states.

\section{Scattering in the helical edge states}\label{sec:scattering}
As we discussed previously, several experimental results already fit nicely the theoretical predictions. (i)~A topological QSHI phase arises in the inverted regime. (ii)~If the system is in the topological regime and if Fermi energy is in the bulk energy gap, current flows only on the edges. (iii)~Transport on the edges occurs via two 1d counter propagating channels, which (iv)~only are weakly sensitive to backscattering in the presence of TRS, leading to ballistic or quasi-ballistic transport over $\mu$m distances. (v)~The edge states are spin-polarised, suggesting that they form helical liquids, where electrons with opposite spin polarisation propagate in opposite directions.

In the presence of TRS, single particle elastic backscattering is prohibited by Kramers theorem. Therefore, current flows ballistically on the edges of 2d TIs, as has already been observed in short samples.
In this regime, the resistance is quantized and independent of the distance between the contacts, in agreement with the red and green curves in Fig.~\ref{fig:konig_quant} and in Fig.~\ref{fig:Du15_T}(a).
However, the resistance starts increasing for longer systems, and departs from the quantized plateau. This is shown both by the blue curve in Fig.~\ref{fig:konig_quant} for HgTe/CdTe QWs and by the different curves in Fig.~\ref{fig:Du15_T}(b) for InAs/GaSb systems, and signals the onset of some backscattering mechanism.
Identifying the leading backscattering mechanisms at the edge of a 2d TI is crucial both from a fundamental point of view and for possible applications.

It is worth pointing out that the prediction of either perfectly quantized ($G=G_0$) or vanishing ($G=0$) conductance, depending on the number of Kramers doublets being either odd or even respectively, can be modified in the presence of electron interactions.
Indeed it is well known that in the presence of interactions, 1d systems show strong differences compared to their higher dimensional counterparts. In two and three dimensions, Landau's Fermi liquid theory usually provides a description of interacting electronic systems in terms of nearly free quasiparticles with renormalized mass. In one dimension, in contrast, this quasiparticle picture breaks down. Indeed, since a particle cannot move without perturbing all the other ones, the eigenstates of the system turn out to be collective excitations.

The most relevant collective low-energy excitations in one dimension are particle-hole excitations, whose bosonic character contrasts sharply with the fermionic quasiparticles of Fermi liquid theory. Interacting 1d systems are instead described by the Tomonaga-Luttinger theory~\cite{tomonaga1950remarks, luttinger1963exactly, giamarchi2003quantum}.
This fact entails a plethora of wholly novel phenomena in 1d interacting systems, the most celebrated ones being spin-charge separation~\cite{haldane1981luttinger, deshpande2008one, Auslaender01042005, lorenz2002evidence, bockrath1999luttinger} and charge fractionalization~\cite{safi1995transport, maslov1995landauer, steinberg2007charge, kamata2014fractionalized}.

Therefore, one can expect that the electron-electron interactions could play a fundamental role in the helical edge states as well.
In the past years, different physical processes giving rise to backscattering have been studied theoretically.
We start this section by introducing the helical Luttinger liquid model, which allows a nonperturbative treatment of interaction effects. Then we review the main mechanisms proposed to explain the increasing edge resistance in the diffusive regime.

\subsection{The helical Luttinger liquid}
Luttinger liquid theory applies to a large range of 1d quantum systems.
This formalism has been successfully applied to describe interacting fermions in carbon nanotubes~\cite{lorenz2002evidence}, semiconductor nanowires~\cite{Auslaender01042005} and quantum Hall edge states~\cite{chang2003chiral}.
The low-energy degrees of freedom of an electronic 1d system are right- and left-moving electrons with momenta close to the two Fermi points, and with up or down spin polarization along a fixed spin quantization axis.
In the presence of electron-electron interactions, the system can be described as a spinful Luttinger liquid containing bosonic charge and spin modes.
By applying a strong external magnetic field it is possible to polarize the electrons and thus to effectively obtain a spinless Luttinger liquid, where the electron spin is frozen but two different propagation directions remain possible.
More exotic 1d systems can also be treated within this approach, the edge states of QH systems being the most famous example. In this case, a single chiral 1d channel appears at the edge, whose spin polarization and chirality are determined by the strong external magnetic field. The QH edge states are therefore characterized as chiral Luttinger liquids~\cite{chang2003chiral}.

The edge states of 2d TIs give rise to a new paradigm in the 1d world: the helical Luttinger liquid~\cite{wu2006helical}. Here, spin and momentum are locked to each other, with spin up and spin down electrons counter-propagating. We consider a single edge, assuming without loss of generality that right-moving electrons are spin-up polarized and vice versa. They are described by the electron field operators $\psi_{\uparrow}(x)\equiv\psi_{R\uparrow}(x)$ and $\psi_{\downarrow}(x)\equiv\psi_{L\downarrow}(x)$.
The free Hamiltonian of a single edge states is
\begin{equation}\label{eq:H0}
H_0=-iv_F\int~dx~\left (\psi_{\uparrow}^{\dagger}\partial_x\psi_{\uparrow}-\psi_{\downarrow}^{\dagger}\partial_x\psi_{\downarrow}\right )
.\end{equation}
Interactions can generate scattering within one branch as well as inter-branch scattering, which are historically labelled as $g_2$ and $g_4$ terms, respectively,
\begin{eqnarray}\label{eq:Hfw}
H_{\mathrm{2}} &=&g_2\int~dx~\psi_{\uparrow}^{\dagger}\psi_{\uparrow}\psi_{\downarrow}^{\dagger}\psi_{\downarrow}\\
H_{\mathrm{4}} &=&\frac{g_4}{2}\int~dx~\left [\psi_{\uparrow}^{\dagger}\psi_{\uparrow}\psi_{\uparrow}^{\dagger}\psi_{\uparrow}+\psi_{\downarrow}^{\dagger}\psi_{\downarrow}\psi_{\downarrow}^{\dagger}\psi_{\downarrow}\right ] \label{eq:Hch}
.\end{eqnarray}
The difficult problem of treating interactions is to some extent exactly solvable in 1d within the bosonization formalism.
It is based on expressing the fermionic field operators as exponentials of bosonic operators. This mapping is exact in one dimension, and allows to translate the difficult fermionic model into a non-interacting bosonic one if the fermionic single-particle spectrum is linear.
The key ingredients are the bosonized versions of the fermionic operators ($\sigma=\uparrow,\downarrow=+,-$)
\begin{equation}
\psi_{\sigma}(x)=\frac{U_{\sigma}}{\sqrt{2\pi a}}e^{i\left [\theta(x)-\sigma\varphi(x)\right ]},
\end{equation}
where $\phi(x)$ and $\theta(x)$ are bosonic fields satisfying the canonical commutation relations $\left [\phi(x),\partial_y\theta(y)\right ]=i\pi\delta(x-y)$, and $a$ is a short-distance cut-off.
Moreover, $U_{\sigma}$ denote so-called Klein factors satisfying fermionic anti-commutations relations. They are necessary for guaranteeing the correct anticommutation relations between $\psi_\uparrow(x)$ and $\psi_\downarrow(x)$, and should be taken onto account when the number of particles of a given species is not conserved (for example during tunneling).
Within the bosonization formalism the electron density reads $\rho_{\sigma}=-(\sigma/2\pi)\partial_x (\sigma \varphi_{\sigma} - \theta)$, so the interaction Hamiltonians~\eqref{eq:Hfw}-\eqref{eq:Hch} become bilinear in the bosonic fields. Crucially, if the fermionic single-particle spectrum is linear, the kinetic energy (\ref{eq:H0}) also becomes bilinear in $\partial_x\theta(x)$ and $\partial_x\varphi(x)$. This makes it possible to express the helical Luttinger liquid Hamiltonian $H_{\mathrm{hLL}}=H_0+H_2+H_4$ as
\begin{equation}\label{eq:Hll}
H_{\mathrm{hLL}}=\frac{v}{2\pi}\int~dx~\left [\frac{1}{K}\left (\partial_x\phi\right )^2+K\left (\partial_x\theta\right )^2\right ]
.\end{equation}
Describing two counter-propagating channels, Eq.~~\eqref{eq:Hll} is formally identical to the Hamiltonian of a \emph{spinless} LL.
Interactions renormalize the sound velocity of the collective excitations velocity
\begin{equation}
v=v_F\sqrt{\left (1+\bar{g}_4+\bar{g}_2\right )\left (1+\bar{g}_4-\bar{g}_2\right )}
\end{equation}
and define the Luttinger parameter
\begin{equation}
K=\sqrt{\frac{1+\bar{g}_4-\bar{g}_2}{1+\bar{g}_4+\bar{g}_2}}
,\end{equation}
with $\bar{g}_i= g_i/(2\pi v_F)$. The Luttinger parameter $K$ is less than one for repulsive interactions, and describes the strength of interactions in the system: $K\sim 1$ for weak interactions ($K=1$ in their absence), while $K\ll 1$ in the strongly interacting regime.
Note that in the presence of chiral interactions only ($g_2=0$), $K=1$ so that the main effect of the $g_4$ term \eqref{eq:Hch} is to renormalize the Fermi velocity.
On the other hand, the $g_2$ term \eqref{eq:Hfw} implies to $K<1$, thus leading to the emergence of Luttinger liquid behaviour.\\

\subsection{Single impurity}
In ordinary LLs (spinful or spinless), it is possible to write down single-particle (1P) BS terms without violating TRS, like $g_{\mathrm{1P}}=g_{\mathrm{1P}}\psi_{L\sigma}^{\dagger}\psi_{R\sigma}+\mathrm{h.c.}$ (with $\sigma$ a redundant variable in the spinless case), induced for instance by a non-magnetic impurity.
This scattering term has the potential to open a gap in the spectrum of the 1d system, leading to a crossover between a conducting to an insulating state.

A powerful tool to shed light on the crossover between these two regimes is represented by the renormalization group (RG) technique.
This approach allows to make qualitative predictions on the relevance of the different terms involved in the model.
It consists in studying how the parameters of the model are modified while zooming to the low energy sector or, analogously, when looking at the coarse-grained theory.
Within the first approach, one starts from the action $S_{\Lambda}[\{g_i\}]$, function of a set of parameters $\ {g_i\ }$ and defined for momenta below a cutoff $\Lambda$. Then, by integrating out a thin momentum shell around the original cutoff, an effective action $S_{\Lambda-d\Lambda}[\{g_i+dg_i\}]$ is obtained, with renormalized parameters and defined with a smaller cutoff; finally one requires invariance under cutoff rescaling, thus demanding $S_{\Lambda-d\Lambda}\left [\{g_i+dg_i\}\right ]=S_{\Lambda}\left [\{g_i\}\right ]$. This procedure gives back the so-called RG equations for the parameters
\begin{equation}
\frac{dg_i}{dl}=\Delta_ig_i,
\end{equation}
where $l$ is used to parametrize the cutoff scaling $\Lambda(l)=\Lambda_0e^{-l}$.
Therefore, as one approaches lower and lower energies the magnitude of the parameter $g_i$ decreases if $\Delta_i<0$ or increases if $\Delta_i>0$.
In the former case the operator is RG relevant, while in the latter it is RG irrelevant.
The RG flow stops as soon as the highest energy scale of the problem is reached, which sets a typical scale $l^*$. If we assume the temperature is the only important energy scale, we can model $l^*=\ln(D/T)$, with $D$ the bulk energy gap; at $T=0$, $l^*\to\infty$ and the parameters $\{g_i\}$ flow either to negligible, if RG irrelevant, or to very large values, if RG relevant. Therefore, the RG procedure represents a powerful tool to identify the most important physical processes of a model at low energies.

In the case of scattering from single impurity, the RG equation for the spinless LL gives~\cite{kane1992transmission}
\begin{equation}\label{eq:RG1p}
\frac{dg_{\mathrm{1P}}}{dl}=(1-K)g_{\mathrm{1P}}
,\end{equation}
and in the case of spinful LLs one has to replace $K\to K_c$, the Luttinger parameter for the charge sector.
Therefore, even in a weakly interacting system with $K\lesssim 1$ the scattering flows to very high values as the temperature is decreased, and the impurity is responsible for a metal to insulator transition.

The situation is very different in a 1D helical systems. A 1P BS process in a helical liquid is given by $g_{\mathrm{1P},B}\psi_{\downarrow}^{\dagger}\psi_{\uparrow}+\mathrm{h.c.}$, and, because of helicity, it must flip the electron spin.
If TRS is broken, for example by a magnetic impurity, the scaling equation of $g_{\mathrm{1P},B}$ is given by Eq.~\eqref{eq:RG1p}, and predicts an insulating state even at very weak interactions.
However, in the presence of TRS such a term is prohibited, $g_{\mathrm{1P},B}=0$, so that the helical liquid is, at this simple level, insensitive to non-magnetic impurities and disorder, in contrast to its spinless and spinful counterparts.
This conclusion is at the basis of the observation of the universal quantized conductance measured in experiments involving short samples~\cite{konig2007quantum}.
Still, different mechanisms preserving TRS can affect the helical edge conductance.
Despite being less relevant, in the RG sense, than 1P elastic BS processes, they can lead to a transition from conducting to insulating behavior for strong enough interactions.
In the following we review different possible processes occurring in the helical liquid in the presence of TRS, as they can explain the deviations from the quantum plateau observed in samples longer than few micrometers.

\subsection{Umklapp interaction}
The interaction terms Eqs.~\eqref{eq:Hfw}-\eqref{eq:Hch} do not represent all possible terms allowed by TRS. In particular, the Umklapp scattering term
\begin{equation}\label{eq:Hum}
H_{\mathrm{U}}=g_{\mathrm{U}}\int~dx~e^{-i4k_F x}\psi_{\uparrow}^{\dagger}\partial_x\psi_{\uparrow}^{\dagger}\psi_{\downarrow}\partial_x\psi_{\downarrow}+\mathrm{h.c.}
\end{equation}
is also allowed.
Due to the oscillating factor it can be important only if the Fermi momentum is at the Dirac point $k_F=0$. If the Fermi momentum is fine tuned in such a way, the bosonized version of Eq.~\eqref{eq:Hum} is
\begin{equation}
H_{\mathrm{U}}=\frac{g_{\mathrm{U}}}{2(\pi a)^2}\int~dx~\cos\left [4\phi(x)\right ].
\end{equation}
Contrary to the forward and chiral interaction terms which do not induce backscattering, the umklapp term scatters two (spin-down) left-moving electrons into two (spin-up) right-moving ones or viceversa.
Therefore it has the potential to open a gap in the spectrum of the hLL, leading to insulating behaviour.
In the regime in which umklapp scattering is dominated by kinetic energy, the system remains gapless and one expects the conductance to be quantized, $G=G_0$, also in the presence of interactions.
On the other hand, even though Eq.~\eqref{eq:Hum} does not explicitly break TRS, a spontaneous TRS breaking can be induced when umklapp scattering dominates.
If $g_{\mathrm{U}}$ is large enough, the interaction umklapp term can dominate over the kinetic part given by Eq.~\eqref{eq:Hll}. Semiclassically, the ground state of the system must minimize $H_{\mathrm{U}}$, therefore pinning the cosine in Eq.~\eqref{eq:Hum} to one of its minima. However,  once $\phi(x)$ is pinned, TRS is spontaneusly broken. This is reflected in the creation of an energy gap that leads to a conducting to insulating transition.
By following the RG steps one finds
\begin{equation}
\frac{dg_{\mathrm{U}}}{dl}=2(1-2K)g_{\mathrm{U}}
.\end{equation}
Therefore umklapp scattering is irrelevant for not too strong interactions $K>\frac{1}{2}$, and cannot spontaneously break TRS at $T=0$.
On the other hand, if $K<\frac{1}{2}$ it becomes relevant, and at $T=0$ the field $\phi$ becomes pinned and TRS is spontaneously broken~\cite{zhang2014time, orth2015non, ziani2015fractional}.
In this case the system becomes gapped, leading to a vanishing conductance $G=0$.

A general comment is in order at this point. As pointed out by Wu \textit{et al.}~\cite{wu2006helical} and by Xu and Moore~\cite{xu2006stability}, no strict topological distinction between 2d TIs with an even and an odd number of Kramers doublets exists in the presence of interactions.
Indeed, we find that for sufficiently strong interactions also the topologically protected $\nu=\text{odd}$ state can be gapped out, leading to $G=0$.
The $\mathbb{Z}_2$ topological distinction is only rigorous in the absence of interactions, but it can fail when considering interacting systems~\cite{santos2015interaction}.

\subsection{Impurity-induced inelastic two-particle backscattering}
We have discussed how the uniform umklapp scattering term can become relevant for $K<\frac{1}{2}$, leading to localization. However, $k_F = 0$ is required, so that in general other mechanisms must be responsible for increasing of the edge resistance for general Fermi energies in the bulk energy gap. In this sense, the role of impurities must be investigated. As discussed, non-magnetic impurities cannot give rise to elastic backscattering. However, Crepin \textit{et al.}~\cite{crepin2012renormalization} showed, by refining a previous work by Str\"{o}m \textit{et al.}~\cite{strom2010edge}, that two-particle (2P) BS can be generated by a non-magnetic impurity in the presence of Rashba spin orbit coupling, which is likely to occur in realistic samples.
Within these assumptions, the Rashba impurity scattering site embedded into the helical liquid is described by $H=H_{\textrm{hLL}}+H_{\mathrm{R}}$, with the first term given in Eq.~\eqref{eq:Hll} and
\begin{equation}\label{eq:HR}
H_{\mathrm{R}}=\pi a v\int~dx~\alpha(x)\left [\left (\partial_x\psi_{\uparrow}^{\dagger}\right )\psi_{\downarrow}-\psi_{\uparrow}^{\dagger}\left (\partial_x\psi_{\downarrow}\right )\right ]+\mathrm{h.c.},
\end{equation}
with $\alpha(x)=\alpha_0\delta(x-x_0)$, $\alpha_0$ being the dimensionless bare scattering amplitude of the point-like impurity located at $x=x_0$.
After bosonizing Eq.~\eqref{eq:HR}, the RG procedure not only renormalizes the bare parameters, but also generates new processes. In particular, a 2P inelastic BS process is generated~\cite{lezmy2012single} at the scattering site
\begin{equation}
H_{\mathrm{2P}}\propto g_{\mathrm{2P}}\cos\left [4\phi(x_0)\right ],
\end{equation}
whose dimensionless strength flows under RG as
\begin{equation}\label{eq:HRflow}
\frac{dg_{\mathrm{2P}}(l)}{dl}=(1-4K)g_{\mathrm{2P}}(l)+\left (1-\frac{1}{K}\right )\left (1-2K\right )\alpha_0(l)^2
.\end{equation}
The initial condition $g_{\mathrm{2P}}(0)=0$ must be taken, since the 2P BS process is not present in the bare Hamiltonian but can only be generated under RG flow.
Crucially, to generate the 2P inelastic scattering both Rashba coupling ($\alpha_0(0)\neq 0$) and electron interactions ($K\neq 1$) are necessary.
By integrating Eq.~\eqref{eq:HRflow} one finds
\begin{equation}\label{eq:Rginteg}
g_{\mathrm{2P}}(l)=\left (1-\frac{1}{K}\right )\alpha_0(0)^2\left [e^{-(4K-1)l}-e^{-2Kl}\right ]
,\end{equation}
which gives the renormalized strength of the 2P inelastic BS process at the scale $l$.
According to the strength of the electron interaction, two different scenarios can emerge.

\subsubsection{Weak coupling regime}
At the end of the RG flow $l\to\infty$, corresponding to the $T\to 0$ limit, the amplitude Eq.~\eqref{eq:Rginteg} flows to zero provided $K>\frac{1}{4}$, and the 2P BS operator, although generated, is actually irrelevant, the helical liquid remaining gapless.
Therefore, at zero temperature the conductance is perfectly quantized, $G=G_0$. At small but finite temperature, it slightly deviates from the plateau, $G=G_0-\delta G$ due to the small but non vanishing contribution from 2P inelastic BS. In particular, because $\delta G\propto g_{\mathrm{2P}}^2$ to lowest order one expects
\begin{equation}\label{eq:scaling_weak}
\delta G\sim\left \{\begin{matrix}
T^{4K} & K>\frac{1}{2}\\
T^{8K-2} & \frac{1}{4}<K<\frac{1}{2}
\end{matrix} \right.,
\end{equation}
because of the competition between the exponents in Eq.~\eqref{eq:Rginteg}.
Therefore, at weak interaction $K\sim 1$ one expects that 2P inealstic processes give a $T^4$ contribution to backscattering. Note that the scaling Eq.~\eqref{eq:scaling_weak} cannot be simply guessed from the bare form of the 2P BS point-like process introduced by Kane and Mele~\cite{kane2005quantum} proportional to $\psi_{\uparrow}^{\dagger}\partial_x\psi_{\uparrow}^{\dagger}\psi_{\downarrow}\partial_x\psi_{\downarrow}+\mathrm{h.c.}$.
The RG analysis of this operator gives back the correct crossover between the relevant and irrelevant regimes at $K=\frac{1}{4}$, but predicts a uniform scaling for $K>\frac{1}{4}$ as $\delta G\sim T^{8K-2}$, which yields a $T^6$ contribution to $\delta G$ in the weakly interacting regime.
On the other hand, by deriving this process from the combined presence of interactions and Rashba impurity, Crepin \textit{et al.} are able to demonstrate that 2P inelastic processes represent a more important source of scattering in the weakly interacting regime, with a lower power law scaling $\delta G\sim T^4$.

\subsubsection{Strong coupling regime}
Although not present in the initial theory, for $K<\frac{1}{4}$ the 2P inelatic BS process is generated by RG flow and drives a metal to insulator transition at zero temperature. Indeed, in the limit $l\to\infty$ and for $K<\frac{1}{4}$ the coupling in Eq.~\eqref{eq:Rginteg} $g_{\mathrm{2P}}\to\infty$.
The scalar field $\phi$ is pinned at $\phi_n(x_0)=(2n+1)\frac{\pi}{4}$, the system is pinched off and two semi-infinite helical liquid disconnected at the Rashba impurity site are created: the conductance vanishes, $G=0$.
At small but finite temperature, the conductance is partially restored by thermal fluctuations.
By integrating out the scalar field away from the scattering site, an effective theory for instanton tunneling can be developed, where tunneling events between adjacent minima $\phi_n$ and $\phi_{n+1}$ are allowed and represent the first order perturbation to the pinched off liquid.
This operator is proportional to the tunneling amplitude $t$, which scales under RG as $\frac{dt}{dl}=\left (1-\frac{1}{4K}\right )t$; it is irrelevant (relevant) for $K<\frac{1}{4}$ ($K>\frac{1}{4}$), consistently with the weak-coupling scenario.
The conductance $G\sim t^2$, so that the temperature scaling is
\begin{equation}
G\sim T^{\frac{1}{2K}-2} \ \ \ \ K<\frac{1}{4}
.\end{equation}
At zero temperature $G=0$, but finite temperature allows instanton tunneling between minima separated by $\Delta\phi=\phi_{n+1}-\phi_n=\frac{\pi}{2}$; from the relation $j=\frac{2}{\pi}\partial_t\phi$, the total charge $\Delta Q=\int~dt~j$ transferred by this process across the scattering site is $e/2$.

In this section we have considered a single point-like Rashba impurity.
In the case of randomly distributed Rashba disorder, Geissler \textit{et al.}~\cite{geissler2014random} have shown that, despite the power law behavior of the conductance is in general changed, the crossover between the weak and the strong coupling regime still occurs for $K=\frac{1}{4}$.

\subsection{Single particle inelastic backscattering}
In the previous section we have shown that the combination of Rashba impurity potential and electron interactions give rise to 2P inelastic BS.
It is natural to wonder if processes involving inelastic scattering of a single electron can be generated.

\subsubsection{Electron-phonon coupling}
One possibility is represented by 1P inelastic BS as brought by electron-phonon coupling in the presence of Rashba spin-orbit coupling.
However, it is possible to show that, when the Rashba potential in Eq.~\eqref{eq:HR} is considered, no correction to the quantized conductance arises. Therefore, to first approximation, the helical edges are robust against 1P inelastic BS generated by electron-phonon coupling.
Beyond terms linear in momentum, TRS allows other terms with higher odd powers of momentum.
These terms can give a contribution to $\delta G$, but in principle they are less relevant in the RG sense.
In particular, the Rashba term cubic in momentum is predicted to be responsible for a weak temperature dependence $\delta G\sim T^6$ at weak interactions~\cite{budich2012phonon}.

\subsubsection{Kondo impurity}
A different mechanism can be generated due to potential inhomogeneities in the vicinity of the edge. Although 1P elastic BS is prohibited, these inhomogeneities can trap bulk electrons, which can in turn interact with edge electrons.
An edge electron can then flip its spin and be backscattered, provided the bulk electron flips its spin as well, so that TRS is preserved.
By considering a single electron trapped near the edge, Maciejko \textit{et al.}~\cite{Maciejko2009kondo} have mapped this problem onto a Kondo model.
Although this process may lead to localization of the edge electrons, by RG arguments they showed that the process is irrelevant, so that the formation of a local Kramers singlet completely screens the impurity spin, effectively removing the impurity site and leading to quantized conductance at zero temperature.
It is worth to note that in a spinful LL the Kondo impurity pinches off the 1d system in two disconnected parts at $T=0$ and the conductance vanishes.
This represents another striking manifestation of the protection enjoyed by the helical liquid, beyond the simple non-interacting $\mathbb{Z}_2$ argument.
Note that, at sufficiently low bias, the conductance remains quantized at every (low) temperature, and not only at $T=0$, as shown by Tanaka \textit {et al.}~\cite{tanaka2011conductance}.

\subsubsection{Charge puddles}
The role of bulk potential inhomogeneities near the edge has been investigated from a different perspective by V\"{a}yrynen \textit{et al.}~\cite{vayrynen2013helical}. Starting from the observation that the combined presence of ionized dopant atoms in the QW structure and gate contacts can give rise to the formation of charge and hole puddles in the bulk of the QW, they studied how the coupling of these puddles with the edge electrons can affect their conductance.
The puddles are considered as quantum dots, where the electrons can tunnel in and out after a dwelling time.
During their dwelling in the dot, electrons interact, so that they can undergo to inelastic scattering.
In this picture, both 1P inelastic and 2P inelastic processes can be generated, that contribute to the deviation from the quantized edge conductance as $\delta G=\delta G_1+\delta G_2$ respectively, with $\delta G_1\sim T^4$ and $\delta G_2\sim T^6$.
Depending of the doping level, different scaling of $\delta G$ is expected.
This theory appears likely to explain several experimental observations, such as the resistance fluctuations in short HgTe/CdTe devices and the persistence of persistence of electron propagation on the edge in highly resistive samples.

\subsubsection{Single electron scattering in the absence of axial spin symmetry}
An additional backscattering mechanism was studied by Schmidt \textit{et al.}~\cite{schmidt2012inelastic} in the generic helical liquid introduced in Sec.~\ref{sec:2dTI}. In this model, spin is not a good quantum number. This picture corresponds to general case in the presence of SIA or BIA: these asymmetries do not break TRS, thus keeping the edge states gapless, but introduce a momentum-dependent spin polarization, see Fig.~\ref{fig:generic}.
In the absence of SIA and BIA, as in the BHZ model, by substituting $\psi_{\sigma}(x)=\sum_ke^{ikx}c_{\sigma,k}$ in the free Hamiltonian Eq.~\eqref{eq:H0} one has
\begin{equation}\label{eq:sz}
H_0=v_F\sum_k k\left (c_{\uparrow,k}^{\dagger}c_{\uparrow,k}-c_{\downarrow,k}^{\dagger}c_{\downarrow,k}\right )
,\end{equation}
the operator $c_{\uparrow(\downarrow),k}$ destroying a right-moving spin-up (left-moving spin-down) electron with momentum $k$.
In the most general case of a generic helical liquid, spin is not a good quantum number, but linearly dispersing gapless edge states still exist, so that the free Hamiltonian reads
\begin{equation}\label{eq:generic}
\tilde{H}_{0}=v_F\sum_k k\left (c_{+,k}^{\dagger}c_{+,k}-c_{-,k}^{\dagger}c_{-,k}\right )
,\end{equation}
the operator $c_{+(-),k}$ destroying a right-moving (left-moving) electron with momentum $k$.
The relation between the operators appearing in Eqs.\eqref{eq:sz}-\eqref{eq:generic} can be written in terms of a $k$-dependent rotation matrix~\cite{schmidt2012inelastic, rod2015spin} $c_{\sigma,k}=\sum_{\alpha=\pm}B_{\sigma}^{\alpha}(k)c_{\alpha,k}$ that because of unitarity condition and TRS can be written as~\cite{orth2013point}
\begin{equation}\label{eq:Bk}
B(k)=\left (\begin{matrix}
\cos\left [(k/k_0)^2\right ] & -\sin\left [(k/k_0)^2\right ] \\
\sin\left [(k/k_0)^2\right ] & \cos\left [(k/k_0)^2\right ]
\end{matrix} \right )
.\end{equation}
In Eq.~\eqref{eq:Bk} $k_0$ parametrizes the momentum scale on which the spin quantization axis rotates.
This effect is usually rather small in heterostructures~\cite{rod2015spin}, so that by focusing around the Dirac point one can safely expand Eq.~\eqref{eq:Bk} for $(k/k_0)^2\ll 1$.
The interacting Hamiltonian $H_{\mathrm{e-e}}=\int dx\int dx'~\rho(x)U(x-x')\rho(x')$, with $\rho_{\sigma}=\psi_{\sigma}^{\dagger}\psi_{\sigma}$, in the case of generic helical liquids becomes
\begin{equation}\label{eq:U(q)}
H_{\mathrm{e-e}}=\frac{1}{L}\sum_{k,k',q}\sum_{\alpha,\alpha ', \beta, \beta '=\pm}U(q)\left [B^{\dagger}_kB_{k-q}\right ]^{\alpha,\beta} \left [B^{\dagger}_{k'}B_{k'-q}\right ]^{\alpha ',\beta '}c_{\alpha,k}^{\dagger}c_{\beta,k-q}^{\dagger}c_{\alpha ',k'}^{\dagger}c_{\beta ',k'+q}
.\end{equation}
Many scattering processes appear in Eq.~\eqref{eq:U(q)}. Among these, the combinations $\alpha+\beta+\alpha '+\beta '=$ odd give rise to 1P BS. For example the term with $\alpha=\alpha '=\beta=-\beta '=+$ is in the form $c_{+,k}^{\dagger}c_{+,k-q}^{\dagger}c_{+,k'}^{\dagger}c_{-,k'+q}$, thus corresponding to BS of a single left-moving electron into a right-moving one, accompanied by the creation of a particle-hole pair.
Crucially, these processes require the off diagonal component of the rotation matrix in Eq.~\eqref{eq:Bk} to be non-vanishing, and therefore disappear in the spin-conserved limit $(k/k_0)^2=0$.

In the clean case, 1P BS is thermally activated away from the charge neutrality point; only at $k_F = 0$ the 1P BS terms can contribute to the conductance and produce a correction $\delta G\sim T^5$. 

In the presence of uncorrelated impurities, on the other hand, scattered particles can exchange momentum with the impurity giving a $\delta G\sim T^4$ contribution also for $k_F\neq 0$~\cite{schmidt2012inelastic,Kainaris2014conductivity}. Hence, the absence of additional non-essential symmetries, like the spin conservation in this case, allows new backscattering processes, which can contribute to the deviations from the quantized conductance.

\subsection{Observation of a helical Luttinger liquid}\label{sec:Du}
Different physical mechanisms are likely to shed light on the nature of scattering in the helical edge state beyond the ballistic regime.
Among these, we have reviewed backscattering induced by (i) single particle scattering in the presence of magnetic impurities, (ii) Umklapp scattering, (iii) impurity induced two-particle inelastic scattering, (iv) single-particle inelastic scattering.
We have discussed how each of these processes gives rise to deviations from the quantized conductance; crucially, we have shown that information about the physical process occurring in the liquid can be recovered from the temperature dependence of the linear conductance.
Therefore, a strong temperature dependence of the edge resistance is expected to be observed in the diffusive transport regime.
Surprisingly, the edge resistance appears very weakly temperature dependent, as shown in Fig.~\ref{fig:Du15_T}(d), in a wide range from $20$ mK to $4$ K. Combining with different experiments~\cite{spanton2014images, du2015robust, gusev2014temperature}, this behaviour persists in a wide temperature range up to tens of Kelvin.
If on one hand these experiments had led to propose different scattering mechanism which could give temperature independent coherence length~\cite{pikulin2014interplay}, on the other hand they have pushed physicists to clarify the energy regimes under investigation.
Indeed, the edge resistance $G=\frac{dI}{dV}$ is in general a function of the different relevant energy scales of the system, namely the temperature $T$ and the bias voltage $V$ at the contacts.
It is worth emphasizing that the scaling $\delta G\sim T^{\alpha}$ reviewed in Sec.~\ref{sec:scattering} are valid in the linear-response regime
\begin{equation}\label{eq:regimeT}
eV \ll k_BT
.\end{equation}
In this regime, the current $I\propto V$ and $G$ corresponds to the linear conductance.
According to the specific geometry, a voltage $V\sim \frac{h}{2e^2}I$ is developed in response to the injected current $I$, so that the condition Eq. \eqref{eq:regimeT} can be reformulated as
\begin{equation}\label{eq:condition}
T[\text{mK}]\gg 150 I[\text{nA}]
.\end{equation}
In the experiment by L. Du \textit{et al.}~\cite{du2015robust}, discussed in Sec.~\ref{sec:exp}, bias currents of the order $I\sim 100$ nA were applied: therefore, according to Eq. \eqref{eq:regimeT}, the temperature dependence of the conductance should be investigated in the regime $T\gg 15$ K. However in this regime bulk effects become important (see Fig.~\ref{fig:Du15_T}(b)) and it is difficult to extract information about edge transport.
The temperature regime investigated in that paper $T\lesssim 10$ K cannot allow to identify the temperature dependence of the edge resistance.

\begin{figure}[!tt]
\centering
\includegraphics[scale=1]{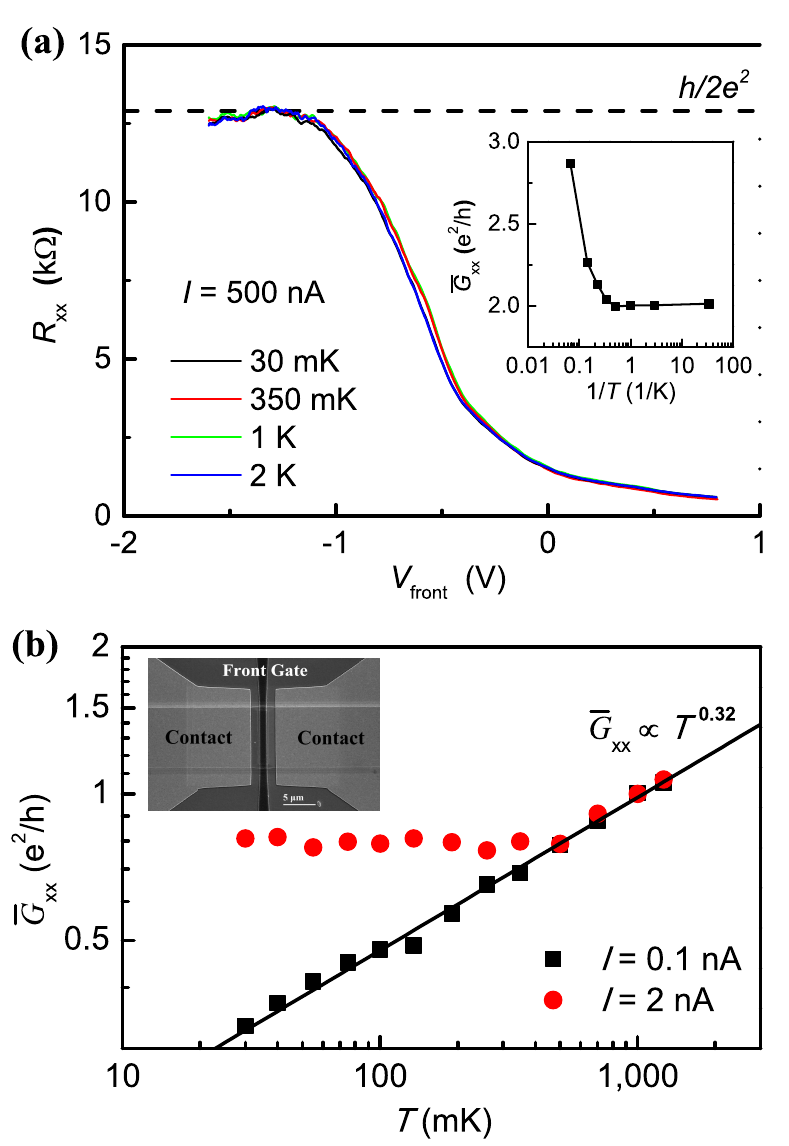}
\caption{(a) Resistance as a function of the gate voltage $V_{\mathrm{front}}$ for different temperatures. The quantized plateau persists from $30$ mK up to $2$ K. For higher temperatures, bulk transport is activated, leading to an increasing of the conductance, as shown in the inset. (b) Log-log plot of the conductance as a function of the temperature for two different applied bias currents. In this temperature range, the bulk contribution is negligible, so that the conductance is safely amenable to the edge. The black straight line represents the power law behaviour $T^{0.32}$. The SEM image of the device is shown in the inset. From Ref.~\cite{li2015observation} with the courtesy of the authors.}
\label{fig:Li_T}
\end{figure}

\begin{figure}[!tt]
\centering
\includegraphics[scale=1]{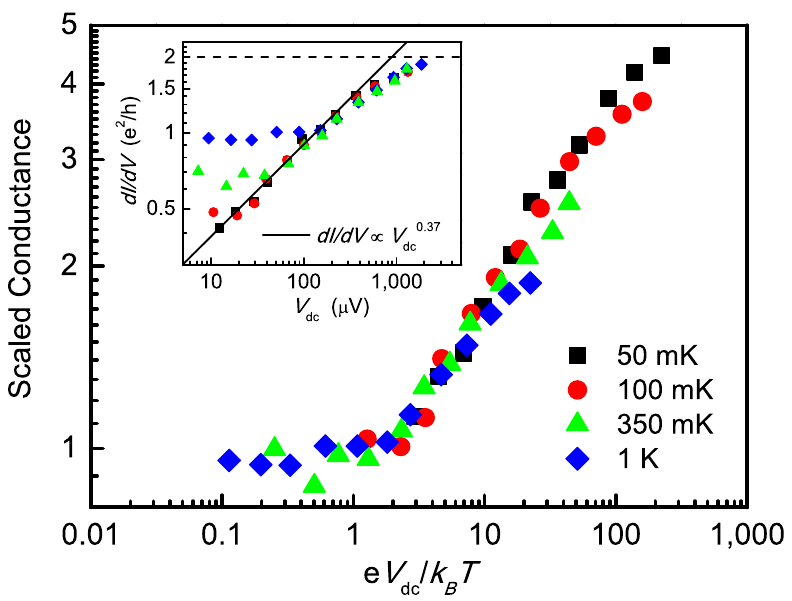}
\caption{Edge conductance as a function of bias voltage measured at different temperatures. In the non-linear regime, a power-law dependence $\frac{dI}{dV_{\mathrm{dc}}}\sim V_{\mathrm{dc}}^{\alpha '}$ is observed, whose fitted exponent is $\alpha '\approx 0.37$. From Ref.~\cite{li2015observation} with the courtesy of the authors.}
\label{fig:Li_V}
\end{figure}

Measuring the temperature dependence of the conductance has proven to be a very difficult issue.
Only recently, the group of Prof. R.-R. Du succeeded in reporting the temperature scaling of the edge conductance in a InAs/GaSb heterostructure at very low temperatures~\cite{li2015observation}.
In this experiment, a main issue is represented by excluding spurious effects, such as non-linear contacts or leaking conductance through the bulk, which could mask or alter the temperature dependence of the edge conductance.
As previously discussed, a temperature independent edge resistance is observed when the condition Eq.~\eqref{eq:condition} is not satisfied, as shown in Fig.~\ref{fig:Li_T}(a).
Here, a current $I=500$ nA is injected, so that, following Eq.~\eqref{eq:condition}, a temperature independent edge conductance is predicted and observed.
To inspect the scattering mechanisms occurring at the edge, the QSH bar must be biased with a very small injected current. Figure~\ref{fig:Li_T}(b) shows the edge conductance as a function of temperature at two different bias currents $I_1=0.1$ nA and $I_2=2$ nA. By following the condition Eq.~\eqref{eq:condition}, two crossover temperatures $T_1\approx 15$ K and $T_2\approx 300$ mK are expected, which separate the $T$-independent from the $T$-dependent regimes.
The experimental behaviour is consistent with Eq.~\eqref{eq:condition}. In particular, the conductance relative to $I_2$ is temperature independent in the non-linear regime $T\ll T_2$, collapsing on the same power law behaviour of the $I_1$ curve for $T\gg T_2$.
Note that the edge conductance scales to zero by lowering the temperature, $G\to 0$ for $T\to 0$.
In the scenario depicted in the previous section, this behaviour is a manifestation of some relevant backscattering process, the behaviour $G\to G_0$ for $T\to 0$ being expected in the irrelevant case. Thus a crucial role must be played by electron interactions, which drive the system to the insulating phase at zero temperature.
Since TRS is preserved in the experiment, a candidate to explain these observation  is represented by two-particle inelastic BS. As shown in the previous section, this mechanism leads to localization for sufficiently strong interactions $K<\frac{1}{4}$. In this regime, at low temperatures non-vanishing conductance is restored by instanton tunneling resulting in $G\sim T^{\frac{1}{2K}-2}$.
The two curves in Fig.~\ref{fig:Li_T}(b) can be fitted with the same power law $G\sim T^{\alpha}$, with $\alpha\sim 0.32$, allowing to extrapolate the Luttinger parameter $K\sim 0.21$. This value is close to the theoretical estimated one $K\sim 0.22$ for InAs/GaSb QWs~\cite{Maciejko2009kondo}, thus supporting the picture of a strongly correlated helical Luttinger liquid.
Note that in the opposite regime $k_BT\ll eV$, non-linear $\frac{dI}{dV}\sim V^{\alpha '}$ characteristics are observed, as shown in Fig.~\ref{fig:Li_V}, confirming the picture of strongly interacting helical edge states with $K\sim 0.21$.

This represents the first manifestation, and at the moment of writing this review the only one, of helical Luttinger liquid behaviour at the edge of a 2d TIs in the QSH phase. Compared with HgTe/CdTe, where the Luttinger parameter is estimated to be $K>0.5$, InAs/GaSb structures have the advantage that in principle the Luttinger parameter can be tuned. Molecular beam epitaxy growth technique and gating architectures allows one to modify the bulk energy gap in these material, in turns modifying the Fermi velocity of the edge state. Since the latter enters the expression of $K$, it should be possible to engineer structures where the role of electron interactions can be controlled and investigated.
In particular, it would be interesting to study the crossover from the insulating to the perfectly conducting edge state at zero temperature, by tuning the Luttinger parameter from the actually measured value $K\sim 0.21<\frac{1}{4}$ to values $K>\frac{1}{4}$, thus allowing to observe the insulating to metal transition predicted to occur at $K=\frac{1}{4}$.

\section{Tunneling dynamics}\label{sec:tunneling}
In Section \ref{sec:scattering} we have discussed the main sources of scattering occurring in the helical liquid, arising from the combined presence of helicity and electron interactions.
If the edges of the QSH bar are well separated, the most relevant scattering processes occur indeed inside the edge, since tunneling from one edge to the other is exponentially suppressed.
On the other hand, if the width of the QSH bar is comparable with the penetration depth of the edge states in the bulk, the wave-functions belonging to different edges can develop a non-vanishing overlap, thus giving rise to a finite tunneling probability~\cite{zhou2008finite, zhang2011electrical}.
In this case, a right-moving electron propagating on one edge can tunnel to the opposite edge, where, due to helicity, it can be backscattered, thus contributing to an increased resistance.
This type of scattering can be minimized by fabricating wide bars: in the micrometer wide bars discussed in Sec.~\ref{sec:exp} tunneling does not play a role in the deviations from the quantized resistance.
Nevertheless, engineering tunneling geometries in the QSH bar can be important in different perspectives.
The quantum point contact (QPC) geometry represents the elementary building block for studying tunneling phenomena between edge states.
This tunneling geometry has been extensively studied, both theoretically and experimentally, in quantum Hall systems.
By creating one or multiple QPC in a fractional QH bar, the fractional charge and statistics of the elementary quasi-particles can be studied, for example by current and noise measurements~\cite{chamon1997two, de1997direct, saminadayar1997observation, radu2008quasi, carrega2011anomalous}.
In the case of 2d TIs, the presence of different spin polarization and chiralities on each edge can make the tunneling dynamics even richer.

Although the experimental realization of QPCs in 2d TIs is still lacking, a variety of theoretical proposals relies on their implementation. The experimental fabrication of tunneling contacts, which is hopefully not too far away, would allow both to study the fundamental properties of the helical edge states, as we discuss in Sec.~\ref{subsec:qpc},  and to develop interesting devices with potential application, some of which are discussed in Sec.~\ref{subsec:interferometry}.

\subsection{The QPC in 2d TIs}\label{subsec:qpc}
\begin{figure}[!tt]
\centering
\includegraphics[scale=0.25]{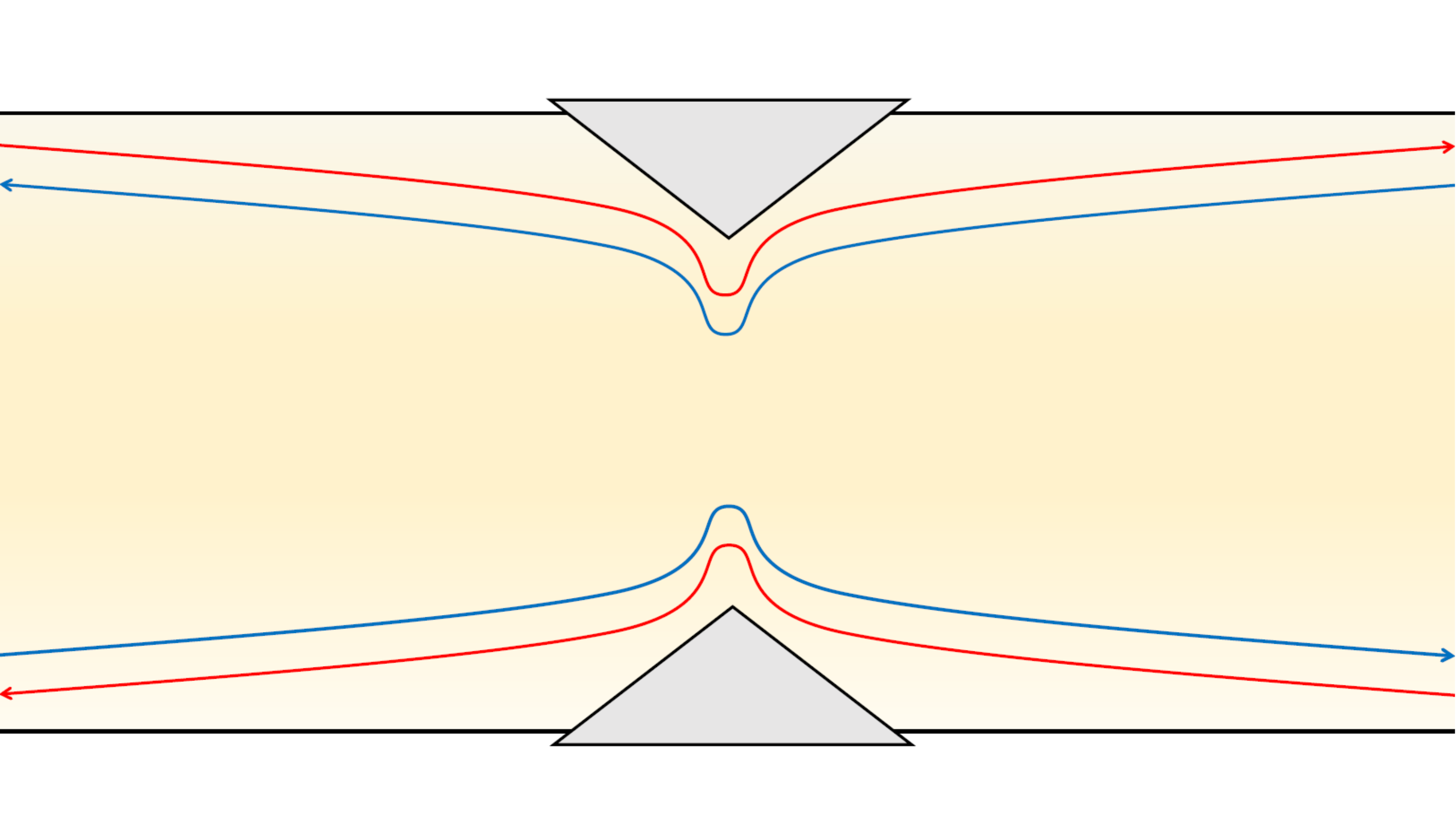}
\caption{Scheme of the quantum point contact realized in the QSH bar. By applying gate voltages (grey regions) the edge states wave-functions can overlap inside the constriction, allowing for tunneling events. Red (blue) lines represent spin-up (spin-down) electrons.}
\label{fig:qpc}
\end{figure}

The QPC geometry is schematically represented in Fig.~\ref{fig:qpc}.
It enables the Kramers doublets on the two edges to be close enough to allow tunneling phenomena from one edge to the other. Experimentally, this setup can be realized either by etching or by gating the edges and manipulating gate voltages in order to create the constriction~\cite{orth2013point, inhofer2013proposal}.
An injected electron approaching the QPC can be either transmitted, remaining on the same edge, or transferred to the other edge.
The conductance of the QSH bar can be written as $G=2G_0-G_{\mathrm{BS}}$, where $2G_0=\frac{2e^2}{h}$ is the quantized conductance of the bar in the absence of tunneling, while $G_{\mathrm{BS}}$ keeps into account the lowering of the conductance due to BS.
In the absence of interactions, the conductance can be evaluated for any transmission amplitude of the QPC via the scattering matrix formalism~\cite{buttiker1992scattering, moskalets2012scattering, sternativo2014tunnel}. However, in general this approach fails when applied to interacting electrons.
Theoretically, problem of evaluating the transport properties of the interacting QSH bar in the presence of a QPC can be approached in two complementary ways: either by taking into account the electron interaction perturbatively and considering the transmission amplitude of the QPC exactly; or by treating the interaction exactly via the Luttinger liquid theory and treating the presence of the QPC perturbatively.
The second approach is particularly suitable if one wants to investigate the role played by interactions.

\begin{figure}[!tt]
\centering
\includegraphics[scale=0.18]{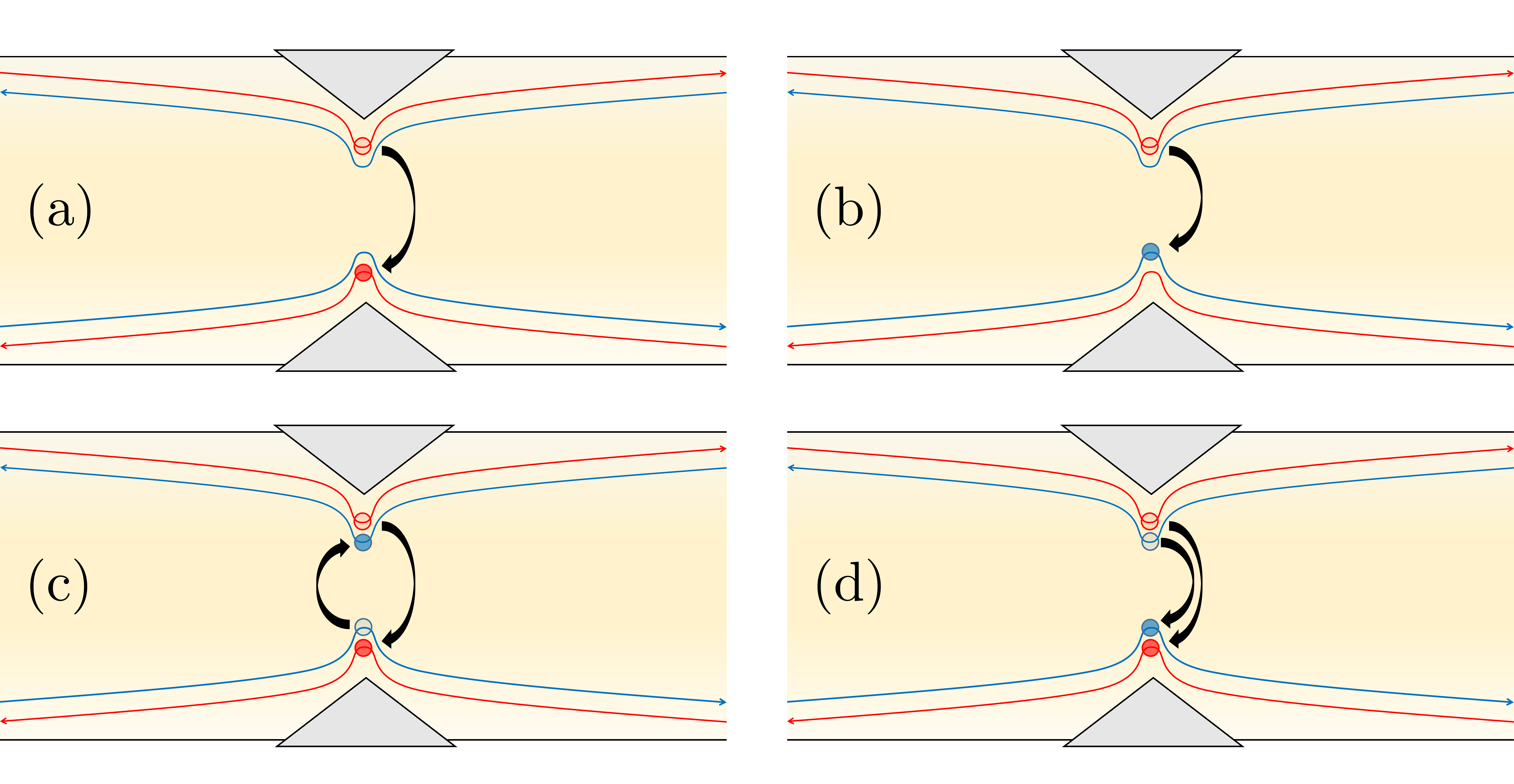}
\caption{(a) Example of 1P SP tunneling, (b) 1P SF tunneling. (c) charge tunneling, which conserves the charge on each edge but not the spin, and (d) spin tunneling, which conserves the spin on each edge but not the charge.
Red (blue) lines represent spin-up (spin-down) electrons.}
\label{fig:processes}
\end{figure}

In the absence of external  magnetic fields, the nature of the tunneling processes is constrained by TRS, which selects the form of the tunneling Hamiltonian $H_{\mathrm{t}}=\int dx~\mathcal{H}_{\mathrm{t}}(x)$.
The term
\begin{equation}\label{eq:H1PSP}
\mathcal{H}_{\mathrm{t,SP}}=r_{\mathrm{SP}}\delta(x-x_0)\sum_{\sigma=\uparrow,\downarrow}e^{i2k_Fx}\psi_{L\sigma}^{\dagger}\psi_{R\sigma}+\mathrm{H.c.}.
\end{equation}
backscatters single electrons through spin-preserving (SP) tunneling events.
In Eq.~\eqref{eq:H1PSP} $r_{\mathrm{SP}}$ represents the tunneling amplitude.
Tunneling is assumed to be point-like, $x_0$ corresponding to the center of the constriction where the overlap between the edge wave functions is maximum.
More realistic models taking into account extended tunnel junctions have been considered~\cite{sternativo2014tunnel, liu2011charge, dolcetto2012tunneling}, but we limit the discussion to the $\delta$-like approximation for simplicity.
Also the spin-flipping (SP) tunneling term
\begin{equation}
\mathcal{H}_{\mathrm{t,SF}}=r_{\mathrm{SF}}\delta(x-x_0)\left [\psi^{\dagger}_{R\downarrow}\psi_{R\uparrow}-\psi^{\dagger}_{L\downarrow}\psi_{L\uparrow}\right ]+\mathrm{H.c.}
\end{equation}
respects TRS; although absent in a strictly spin-conserving model, this process could be generated in the presence of SIA which, even if absent in the QW Hamiltonian, can be induced by the gate voltages used to create the constriction~\cite{vayrynen2011electrical}.
The 1P SP and 1P SF processes are schematically shown in Fig. \ref{fig:processes}(a)-(b) respectively.
The RG equations for their amplitudes read
\begin{equation}
\frac{dr_{\mathrm{SP/SF}}}{dl}=\left (1-\frac{K}{2}-\frac{1}{2K}\right )r_{\mathrm{SP/SF}},
\end{equation}
indicating that 1P tunneling phenomena always represent an irrelevant perturbation to the perfectly conducting QSH bar, regardless the interaction strength. Nevertheless, at finite temperature a correction $G_{\mathrm{BS}}\sim T^{K+\frac{1}{K}-2}$ is predicted~\cite{dolcetto2012tunneling, strom2009tunneling}, which, if measured, could allow to extract the strength of the interaction through the Luttinger parameter $K$.

Furthermore, the interplay between helicity and electron interaction allows peculiar 2P tunneling processes which can dominate over 1P processes.
Indeed, RG flow equations for the so-called charge tunneling process
\begin{equation}\label{eq:Hc}
\mathcal{H}_{\mathrm{C}}=r_{\mathrm{C}}\delta(x-x_0)\psi^{\dagger}_{L\uparrow}\psi^{\dagger}_{L\downarrow}\psi_{R\downarrow}\psi_{R\uparrow}+\mathrm{H.c.}
,\end{equation}
represented in Fig.~\ref{fig:processes}(c), and for the so-called spin tunneling process
\begin{equation}\label{eq:Hs}
\mathcal{H}_{\mathrm{S}}=r_{\mathrm{S}}\delta(x-x_0)\psi^{\dagger}_{L\uparrow}\psi^{\dagger}_{R\downarrow}\psi_{L\downarrow}\psi_{R\uparrow}+\mathrm{H.c.}
,\end{equation}
schematically represented in Fig. \ref{fig:processes}(d), are given by
\begin{eqnarray}
\frac{dr_{\mathrm{C}}}{dl}&=&\left (1-2K\right )r_{\mathrm{C}},\\
\frac{dr_{\mathrm{S}}}{dl}&=&\left (1-\frac{2}{K}\right )r_{\mathrm{S}},
\end{eqnarray}
respectively.
Note that although the spin process is always irrelevant for repulsive interactions $K<1$, the charge tunneling operator can become relevant for strong enough interactions $K<\frac{1}{2}$. In this regime, the system flows to the strong pinch-off limit becoming insulating at zero temperature~\cite{lee2012nonequilibrium}. At small temperature, finite conductance is restored by 2P tunneling processes across the QPC in the strong pinch-off regime~\cite{teo2009critical}, and the conductance displays a new power-law dependence on temperature.

We conclude that the QPC geometry plays an important role in the investigation of the interaction effects in the 2d TIs, displaying a variety of transport regimes as a function of the interaction strength~\cite{teo2009critical, hou2009corner}.
In particular, the strength of electron interactions can be extracted by studying the behaviour of the backscattering current as a function of bias voltage and temperature~\cite{dolcetto2012tunneling, strom2009tunneling, schmidt2011current}, and further information about the interplay between 1P and 2P tunneling processes could be accessed by means of noise measurements~\cite{souquet2012finite, basset2010emission, chamon1995tunneling}.\\

\subsection{Interferometry}\label{subsec:interferometry}
\begin{figure}[!tt]
\centering
\includegraphics[scale=0.25]{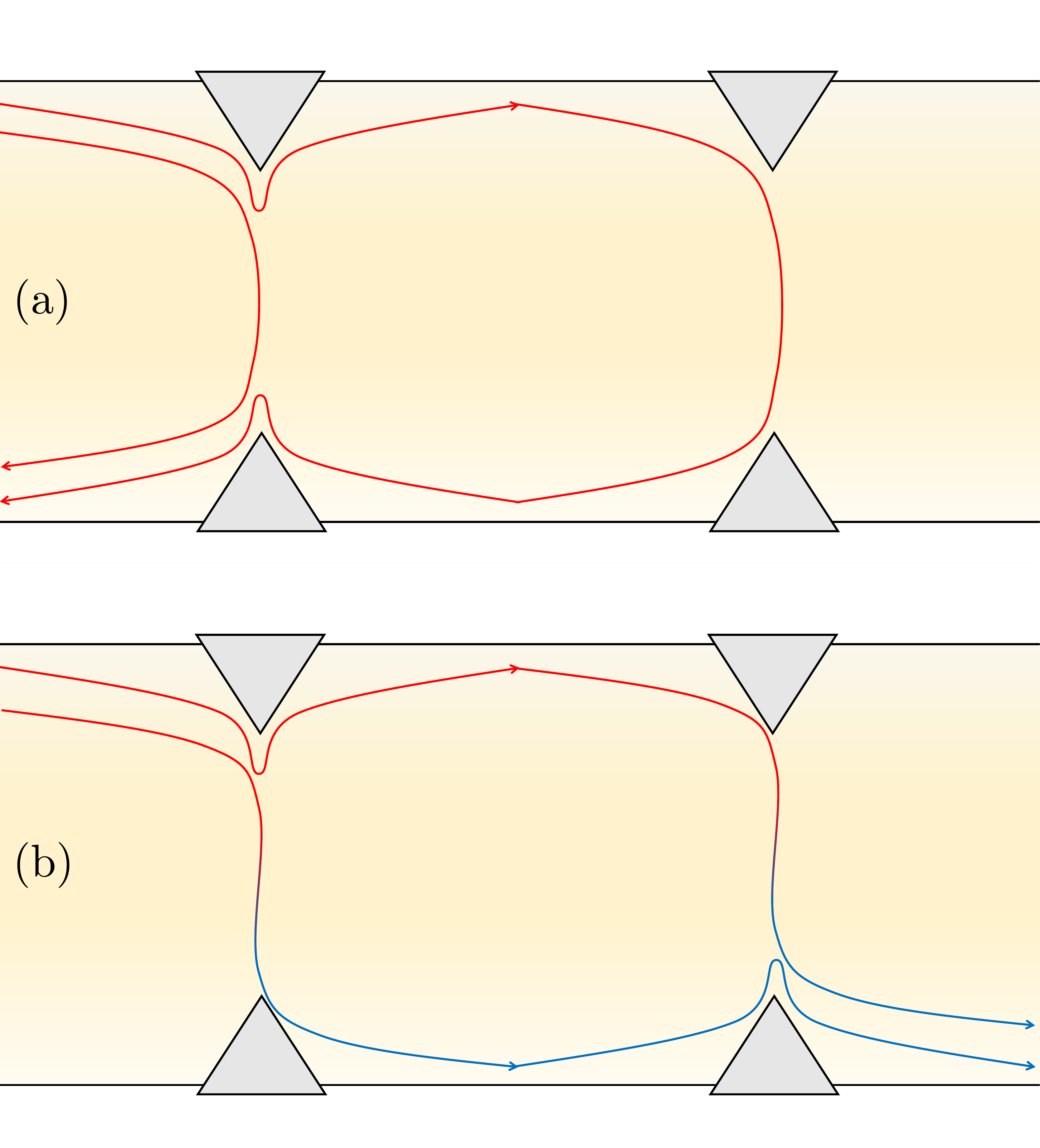}
\caption{Interferometric device realized in a 2d TI based on the helical edge states. Two QPCs split the incoming flux into transmitted and reflected components. The outgoing currents result from quantum interference between different paths performed inside the interferometer. By acting with gate voltages or magnetic fluxes it is possible to manipulate the dynamical phases acquired by the different paths, thus effectively controlling the interference pattern.
The quantum interference can arise from both SP (a) and SF (b) tunneling events.
Red (blue) lines represent spin-up (spin-down) electrons.}
\label{fig:interference}
\end{figure}

The experimental realization of QPCs in 2d TIs would not only represent an interesting playground to test theoretical predictions and to explore the physics of the interacting edge states, but would also allow to implement a variety of interesting applications.
The ability of creating tunneling junctions between the helical edge states can allow one to manipulate their transport properties in a controlled manner. An injected electron can be selectively split into transmitted and reflected channels~\cite{zhang2011electrical, krueckl2011switching, dolcetto2013generating, dolcini2015noise}; together with their intrinsic spin polarization, the realization of such a geometry can pave the way for engineering interesting spintronic devices, where the charge and spin flows could be generated and controlled~\cite{datta1990electronic}.\\
The edge states of 2d TIs also represent an intriguing platform for studying electronic quantum interference.
While optical interferometers are well established since a long time ago, exploring the wave-like nature of electrons has proven a very difficult task.
Due to decoherence experienced by electrons in macroscopic systems, only mesoscopic devices have the potential to shed light on this quantum mechanical effect. In this context, 1d channels, such as the ones appearing at the border of QH or QSH insulators, represent the electronic version of optical fibres: edge electrons can propagate ballistically over $\mu$m distances and, being only very weakly affected by backscattering, preserve their quantum coherence. Analogously, the QPC acts as the electronic counterpart of optical beam splitters,  with the transmitted and reflected components propagating along different paths.
Protected edge channels and tunneling regions thus represent the building blocks for implementing electron quantum interference experiments.
Remarkably, several important results in this direction have already been obtained in integer QH based devices~\cite{grenier2011electron, bocquillon2013coherence, bocquillon2014electron, wahl2014interactions, jullien2014quantum}.\\
The presence of counter-propagating spin-polarized electrons can make the interferometric properties of the helical edge states even richer than the ones emerging in QH chiral edge states~\cite{Chu2009coherent, dolcini2011full}. The setup schematically shown in Fig.~\ref{fig:interference} represents a possible electronic interferometer realized in a QSH bar by means of two QPCs~\cite{dolcini2011full, citro2011electrically, romeo2012electrical, ferraro2013spin}.
Each QPC acts as a beam splitter for incoming electrons. The electron flow incoming into the left QPC is partially transmitted, remaining on the same edge, and partially reflected to the other edge, either preserving or flipping its spin, as show in in Fig.~\ref{fig:interference}(a) and (b) respectively. Analogous processes arise at the right QPC, and the currents collected at the different contacts depend on the recombination of electron fluxes which have performed different paths in the setup.
The interference patterns depend on the different dynamical phases accumulated by the electron while propagating in the different arms of the device. By acting with magnetic fluxes or gate voltages, it is possible to control the dynamical phases acquired by the different paths, thus making it possible to control the quantum interference.
It is worth to note that electron interferometers are not only interesting for investigating the fundamental properties of quantum physics in solid state devices, but could also be exploited to generate and manipulate charge and spin transport in a controlled way.
Furthermore, in the presence of interactions, a richer scenario is expected to emerge, due to the possible interference of fractionalized excitations with bosonic character~\cite{calzona2015time, calzona2015transient}, analogously to what has already been predicted~\cite{wahl2014interactions} and observed~\cite{kamata2014fractionalized, bocquillon2013coherence} in QH devices.

Finally, a step forward towards the implementation of electron quantum optics is represented by the realization of single particle sources (SPSs), which enable to study the quantum nature of single electrons. SPSs have been recently achieved for injecting single electrons and holes into integer QH edge states, either by means of driven mesoscopic capacitors~\cite{moskalets2012scattering, feve2007demand, mahe2010current, buttiker1993mesoscopic} or by designing Lorentzian-like voltage pulses~\cite{dubois2013integer, dubois2013minimal, jullien2014quantum}. With this technology it is possible to investigate quantum interference at the single electron level~\cite{bocquillon2013coherence, wahl2014interactions}.
Analogously, different schemes for creating SPSs in 2d TIs have been proposed~\cite{inhofer2013proposal, hofer2013emission, ferraro2014electronic}. Combined with the experimental realization and precise characterization of QPCs, these architectures would pave the way to study electron quantum optics in topological insulator based devices.

\section{Outlook}\label{sec:outlook}
Almost a decade after their first theoretical prediction and experimental realization, the interest in topological insulators still seems to continue.
Their intriguing fundamental properties, as well as their huge potential for experimental applications, make them extremely fascinating for a large scientific community.
A leading role in this perspective is played by their edge states.
These are characterized by helicity, which binds together momentum and spin-polarization. In two dimensions, this property implies the existence of one-dimensional channels, where electrons with opposite spins counter-propagate.
Due to their one-dimensional nature, electron interactions play a crucial role in their physical properties, giving rise to a new paradigmatic state of matter, called helical Luttinger liquid, which was recently experimentally observed.
Their topological nature considerably limits backscattering, allowing for ballistic propagation in mesoscopic devices.
The weak sensitivity to phase-breaking perturbation could enable to implement intriguing devices, where quantum effects play a prominent role. Although it is yet to come, the experimental realization of the fundamental building blocks of electron quantum optics, such as quantum point contacts and single electron emitters, would allow not only to investigate the fundamental physics underlying these systems, but also to develop intriguing devices with potential applications.

\acknowledgments
G.~D. and T.~L.~S. are supported by the National Research Fund, Luxembourg under grant ATTRACT 7556175. G.~D. and M.~S. acknowledge the support of CNR-SPIN via Seed Project PGESE003.

\end{document}